%% file: TETC-2025-03-0122.R2_Imram.tex
\documentclass[journal]{IEEEtran}
    \usepackage{amsmath,amsfonts}
     \usepackage{algorithm}
     \usepackage{algorithmic}
     
     \usepackage{array}
     \usepackage[caption=false,font=normalsize,labelfont=sf,textfont=sf]{subfig}
     \usepackage{placeins}  % Add this in the preamble
     \usepackage{textcomp}
     \usepackage{stfloats}
     \usepackage{url}
     \usepackage{verbatim}
     \usepackage{graphicx}
     \usepackage{cite}
     \usepackage{xcolor}
     \usepackage{orcidlink}
     \usepackage{pifont}
     \usepackage{empheq}
     \usepackage{eqparbox}
     \usepackage{multicol}
     \usepackage{multirow}
    \usepackage{amssymb}
    \usepackage{array}
    \usepackage{mathtools}
    \usepackage[table]{xcolor}% http://ctan.org/pkg/xcolor
    \usepackage{pgfplots} % package used to implement the plot  
    \pgfplotsset{width=6.5cm, compat=1.5} 
    \usepackage{color,soul}
    \usepackage{comment}
    \newcommand{\cmark}{\textcolor{blue}{\ding{51}}}

    \usepackage{caption}

    \newcommand{\malik}[1]{{\color{black}#1}}
    \newcommand{\muhammad}[1]{{\color{black}#1}}

\begin{document}

    \title{\malik{PIP-NTT: Towards a Scalable Memory-Parallelized Accelerator for Iterative NTT in PQC}}

    %%=====================================
    %% Authors
    %%=====================================
    \author{
        Malik~Imran~\orcidlink{0000-0002-1900-6387}, %~\IEEEmembership{Member,~IEEE,}         
        Ayesha~Khalid~\orcidlink{0000-0002-4815-6966},
        Ciara~Rafferty~\orcidlink{0000-0002-3670-366X},
        Safiullah~Khan~\orcidlink{https://orcid.org/0000-0001-8342-6928},
        Muhammad~Rashid~\orcidlink{0000-0001-5852-1296} and
        ~Máire~O'Neill~\orcidlink{0000-0002-6865-6212}%~\IEEEmembership{Member,~IEEE}% <-this % stops a space

        % \thanks{This work is supported from the Integrated Quantum Networks (IQN) Research  Hub (EP/Z533208/1).}

        % \thanks{This work is funded by the grant from the Engineering and Physical Sciences Research Council (EPSRC) Quantum Communications Hub (EP/T001011/1).

        \thanks{
        % Manuscript received March xx, 2025; revised xx xx, 2025. 
        This work is funded by the Integrated Quantum Networks (IQN) Research  Hub (EP/Z533208/1).\textit{(Corresponding Author: Malik Imran)}
            
            Malik Imran, Ayesha Khalid, Ciara Rafferty and Maire O'Neill are with the Center for Secure Information Technologies (CSIT), Queen's University, Belfast, Northern Ireland, UK. (e-mail: \{m.imran@qub, a.khalid@qub, c.m.rafferty@qub, m.oneill@ecit.qub\}.ac.uk)

            Safiullah Khan is with the Department of Computing and Mathematics, Manchester Metropolitan University, Manchester, UK. (e-mail: safiullah.khan@mmu.ac.uk)
    
            Muhammad Rashid is with the Computer and Network Engineering Department, Umm Al-Qura University, Makkah, Saudi Arabia. (e-mail: mfelahi@uqu.edu.sa)
        }     
        } 

    \markboth{
    Accepted in IEEE Transactions on Emerging Topics in Computing (TETC), 2026. Final version will appear on IEEE Xplore.
    % IEEE Transactions on Emerging Topics in Computing,~Vol.~XX, No.~X, July~2026
    }%
     {M. Imran \MakeLowercase{\textit{et al.}}: PIP-NTT: Towards an Efficient Memory-Parallelized NTT Accelerator}


\maketitle
\thispagestyle{empty}

    \begin{abstract}
        
       \malik{The iterative forward and inverse number theoretic transform (NTT) is a key component in lattice-based post-quantum cryptography (PQC), typically implemented using Cooley–Tukey and Gentleman–Sande butterfly units.     Existing iterative NTT accelerators often rely on ping-pong memory schemes and large memory blocks tied to the cyclotomic ring, which limits overall efficiency. To overcome this, we propose a memory-parallelization strategy using four smaller $\frac{n}{4}$-sized memories for ring size $n$, preserving the total memory footprint of conventional designs. We also introduce a multiplication-free rescaling architecture for the inverse NTT.  Building on these innovations, we perform a comprehensive hardware-based design space exploration of unified Cooley–Tukey and Gentleman–Sande butterfly units, evaluating both coarse- and fine-grained pipelining strategies. The resulting optimized butterfly unit forms the core of our proposed pipelined and memory-parallelized NTT accelerator, ``PIP-NTT''. It integrates two such units alongside the memory-parallelization scheme to boost computational throughput under tight area constraints. Experimental results on FPGA platforms show that PIP-NTT achieves 2.67$\times$ and 1.48$\times$ higher efficiency in average Area-Time Product compared to the most area-optimized and high-speed NTT accelerators in the literature. The design is scalable across butterfly radices and adaptable to other PQC schemes, making it a versatile solution for future cryptographic hardware.}
        \end{abstract}


    %%=============================================%%
    %% \textcolor{blue}{Section}: Introduction
    %%=============================================%%

    \section{Introduction}

    \IEEEPARstart{L}{attice-based} cryptographic accelerators are vital for a wide range of applications, including secure communication in embedded systems, high-performance computing in data centers, and secure Internet of Things (IoT) networks~\cite{crypto-applications}. While companies such as Intel and IBM have developed high-throughput accelerators for classical algorithms like elliptic curve cryptography (ECC) and Rivest–Shamir–Adleman (RSA)~\cite{Intel_QuickAssist_Technology, IBM}, these schemes rely on mathematical problems (such as discrete logarithms and prime factorization) that can be efficiently solved by quantum computers using Shor’s algorithm~\cite{Shors_Algorithm}. This vulnerability has accelerated the global shift toward post-quantum cryptography (PQC)~\cite{tcas_II_malik}.

    In response, the National Institute of Standards and Technology (NIST) officially standardized ML-KEM~\cite{CRYSTALS_Kyber_site} and ML-DSA~\cite{Dilithium} in 2024 as part of its PQC portfolio. Both schemes rely on polynomial multiplication, which is recommended to be performed using the number theoretic transform (NTT) for efficiency~\cite{ntt_comparison}. The NTT comprises forward (FNTT) and inverse (INTT) transforms, whose execution speed directly impacts the overall performance of ML-KEM and ML-DSA cryptosystems~\cite{COHA_NTT, KALI, ntt_kyber_16BTF, pipeline_ntt, ntt_kyber_2, high_speed_NTT_kyber_2024}.

    \malik{To enable practical deployment of ML-KEM and ML-DSA across systems ranging from embedded devices to large-scale cloud infrastructures, efficient hardware acceleration of NTT is essential~\cite{crypto-applications, tcas_II_malik}. NTT accelerators must optimize both throughput and area to meet diverse performance and resource constraints. High-throughput architectures typically employ deep pipelining and parallel memory access~\cite{ntt_kyber_16BTF, HyperNTT, K_NTT_64_BTFs_2024}, but often incur substantial resource overhead, limiting their suitability for constrained platforms. Conversely, area-optimized designs~\cite{ntt_kyber_1, safi_ntt, ESL_malik, sajjad_access_ntt, K_NTT_only_AREA_2024} reduce hardware footprint and simplify control logic, but may suffer from performance bottlenecks. Therefore, a balanced strategy, jointly optimizing throughput and area, is critical for building scalable, energy-efficient NTT accelerators that support diverse PQC deployment scenarios.}

    %%=============================================%%
    %% Related Work
    %%=============================================%%

    \subsection{Related Work}\label{related}

     State-of-the-art NTT architectures are commonly implemented on FPGAs and ASICs using Cooley–Tukey (CT-BU) and Gentleman–Sande (GS-BU) butterfly units~\cite{ntt_comparison, COHA_NTT, KALI, ntt_kyber_16BTF, pipeline_ntt, ntt_kyber_2, high_speed_NTT_kyber_2024, unif_ntt_K_D_2024, ntt_k_plantard, CRYPHTOR, SCA_on_NTT_2024, split_radix_based_K_2024, HyperNTT, poly_mult_acc_for_K_2024, K_NTT_64_BTFs_2024, ntt_kyber_1, safi_ntt, ESL_malik, sajjad_access_ntt, ntt_radix4_2024, Ziying_2023, AREA_TIME_ML_KEM_4BTU_2025, K_NTT_only_AREA_2024}. In~\cite{ntt_comparison}, resource-sharing techniques integrate modular multiplication, reduction, and addition/subtraction units with pipelining to optimize the critical path. Similarly,~\cite{COHA_NTT} presents a configurable NTT design supporting diverse PQC parameters. A notable design for Kyber and Dilithium is introduced in~\cite{KALI}, featuring a 23-bit polynomial multiplier that computes one coefficient per cycle for Dilithium and two for Kyber.

     While~\cite{ntt_comparison, COHA_NTT, KALI} emphasizes scalable NTT designs, parallel architectures for clock cycle optimization are explored in~\cite{ntt_kyber_16BTF, pipeline_ntt, ntt_kyber_2, high_speed_NTT_kyber_2024}. In~\cite{ntt_kyber_16BTF}, a modular multiplier, reduction unit, adder/subtractor, and two ``DIVby2'' blocks eliminate the need for multiplication by $n^{-1}~\texttt{mod q}$. The work in ~\cite{pipeline_ntt} employs multiple modular operators, including Montgomery-based reduction blocks. The FPGA-based design in~\cite{ntt_kyber_2} uses digital signal processor (DSP) blocks and Gentleman–Sande butterflies for both FNTT and INTT, reducing clock cycles by 29\% for Kyber-512 and 33\% for Dilithium-1024 over the schoolbook method. In~\cite{high_speed_NTT_kyber_2024}, clock cycles are optimized using 10 block RAMs (BRAMs).

    In addition to scalable and parallel/pipelined designs, high-speed NTT hardware accelerators are explored in~\cite{unif_ntt_K_D_2024, ntt_k_plantard, CRYPHTOR, SCA_on_NTT_2024, split_radix_based_K_2024, HyperNTT, poly_mult_acc_for_K_2024, K_NTT_64_BTFs_2024}. A unified high-speed NTT accelerator for Kyber and Dilithium is presented in~\cite{unif_ntt_K_D_2024}, while~\cite{ntt_k_plantard} improves Plantard arithmetic for Kyber on low-end IoT devices. Memory-unified designs are proposed in~\cite{CRYPHTOR}, and side-channel vulnerabilities are analyzed in~\cite{SCA_on_NTT_2024}. Split-radix acceleration~\cite{split_radix_based_K_2024}, multi-level pipelining with optimized reduction~\cite{HyperNTT}, and reconfigurable polynomial multipliers~\cite{poly_mult_acc_for_K_2024} further enhance Kyber performance. A 64-butterfly-unit high-speed architecture is demonstrated in~\cite{K_NTT_64_BTFs_2024}.

    \malik{Beyond throughput-oriented designs, several works explore memory organization strategies to balance area efficiency and control simplicity~\cite{ntt_kyber_1, safi_ntt, ESL_malik, sajjad_access_ntt, ntt_radix4_2024}}. A ping-pong memory scheme is used in~\cite{ntt_kyber_1} for Kyber, employing two BRAMs, one block ROM (BROM), and two 5-stage pipelined butterfly units. Similar schemes are adopted in~\cite{safi_ntt, ESL_malik, sajjad_access_ntt, ntt_radix4_2024}. The work in ~\cite{safi_ntt} proposes an FNTT architecture with Barrett-based reduction, while~\cite{ESL_malik} evaluates three Kyber-based NTT designs: FNTT only, INTT only, and a unified design.  Their energy and power evaluations reveal that the unified design is more energy-efficient, while dedicated designs offer better power efficiency on FPGA and ASIC platforms. In~\cite{sajjad_access_ntt}, a unified NTT accelerator shares arithmetic units across CT-BU and GS-BU. A configurable radix-4 design with 10 BRAMs and parallel ping-pong control is presented in~\cite{ntt_radix4_2024}. An efficient BRAM-free core is demonstrated in~\cite{Ziying_2023}, and~\cite{AREA_TIME_ML_KEM_4BTU_2025} explores four butterfly units for area-time trade-offs. Area-optimized NTT design is discussed in~\cite{K_NTT_only_AREA_2024}.

    \malik{
    The designs discussed above demonstrate that while butterfly pipelining is widely adopted across many NTT accelerators~\cite{ntt_comparison, pipeline_ntt, KALI, HyperNTT, ntt_kyber_1, iter_ntt_2025_wiley}, a Kyber-specific pipelined design space exploration within iterative and area-constrained settings has not been systematically addressed in prior work. Furthermore, many high-performance designs attempt to reduce memory access overhead by scaling hardware resources, such as deploying multiple butterfly units and large memory arrays~\cite{COHA_NTT}, which is not practical for resource-limited platforms or scalable across diverse PQC schemes. In contrast, area-constrained designs often adopt ping-pong memory access schemes~\cite{ntt_kyber_1, safi_ntt, ESL_malik, sajjad_access_ntt, ntt_radix4_2024}, which simplify control logic but inherently limit scalability, parallelism, and frequency optimization. These limitations motivate our proposed memory-parallelization strategy, which enables partial parallelism without increasing the total memory cost. Combined with a unified pipelined butterfly architecture, our approach supports scalable deployment across different ring sizes and modular arithmetic configurations, making it well-suited for Kyber and adaptable to other PQC algorithms.
    }
   
    %%=============================================%%
    %% Contributions
    %%=============================================%% 

    \subsection{Contributions}\label{subsec:contributions}

    \malik{Motivated by the limitations highlighted in~Section~\ref{related}, we target iterative NTT architectures for area-constrained applications, aiming to improve performance without increasing hardware cost. To this end, we present a pipelined and memory-parallel NTT accelerator that reduces both clock cycles and resource usage. The proposed design is implemented using Kyber parameters ($n=256$, $\texttt{q}=3329$) on FPGA platforms. Our contributions, in a bottom-up approach, are as follows:}

    \begin{itemize}
        \item [(i)]  
        
        \malik{We perform a comprehensive design space exploration (DSE) to develop a unified and deeply pipelined butterfly unit architecture for Cooley-Tukey and Gentleman-Sande configurations. The DSE evaluates eight architectural variants, including one non-pipelined baseline and seven pipelined designs, using coarse-grained\footnote{Coarse-grained pipelining inserts registers between major functional blocks such as multiplexers, arithmetic units, and modular reduction stages.} and fine-grained\footnote{Fine-grained pipelining adds further registers within modular reduction units to improve timing.} pipelining strategies. The optimized butterfly unit achieves a frequency improvement from 46 MHz to 298 MHz while maintaining minimal area overhead. Detailed results are provided in Section~\ref{sec:dse}.}

        \item[(ii)] 
        
        \malik{We propose a novel multiplication-free architecture for the rescaling step required after the INTT. For Kyber’s modulus $\texttt{q}=3329$, we exploit the congruence $n^{-1} \equiv -13~(\texttt{mod}~3329)$ to compute $a \cdot n^{-1}~\texttt{mod q}$ as $-13 \cdot a~\texttt{mod}~3329$. This is implemented using modular doublings and additions to compute $13a$, followed by a conditional negation. Compared to a 7-term shift–add circuit, our design reduces slice and look-up table (LUTs) usage by 1.67$\times$ and 1.84$\times$, respectively, with only a 1.17$\times$ increase in flip-flops (FFs) due to local buffering. Implementation details are in Section~\ref{sec:proposed_architecture}.}

        \item[(iii)] 
      
        \malik{We introduce a memory-parallelization technique that enables partial parallelism in fully iterative NTT designs. It reduces the number of clock cycles by half while maintaining the same total memory as traditional ping pong schemes. The method stores coefficients in adjacent pairs and uses a multiplexer-based, offset-controlled swap network to reorder outputs from four smaller memory blocks of size $\frac{n}{4}$. This allows two butterfly units to operate concurrently in each cycle. Further architectural details are provided in Section~\ref{sec:proposed_memory_parallel_approach}.}

        \item[(iv)] 
        \malik{We present a complete pipelined and memory-parallel NTT accelerator architecture, named ``PIP-NTT'', which integrates two instances of the optimized butterfly unit and the proposed memory-parallelization strategy. The architecture supports both FNTT and INTT operations through a unified control flow and is implemented on an FPGA using Kyber parameters. As detailed in Sections~\ref{sec:proposed_architecture} and~\ref{sec:results_and_comparisons}, PIP-NTT achieves a 200 MHz operating frequency and computes one FNTT or INTT in 2.60$\mu$s.}
    \end{itemize}

    \malik{The proposed PIP-NTT is comprehensively evaluated against recent state-of-the-art NTT accelerators. It achieves 2.67$\times$ and 1.48$\times$ higher efficiency in average Area-Time Product (ATP) compared to the most area-optimized and high-speed designs in the literature, respectively. Furthermore, the proposed architecture is scalable across different radices and adaptable to other PQC schemes, such as Dilithium, making it a versatile and resource-efficient solution for future post-quantum cryptographic hardware designs.}
    
    %%=============================================%%
    %% Section 2: Background 
    %%=============================================%%

    \section{Mathematical Background}\label{sec:background}

    The \text{FNTT} and \text{INTT} of an $n$-degree polynomial $a(x)$ can be defined by Eq.~\ref{eq:FNTT} and Eq.~\ref{eq:INTT}, respectively.

    \vspace{-2.5mm}
    \begin{equation}\label{eq:FNTT}
        \text{FNTT} = \tilde{a}_i = \sum_{j=0}^{n-1} a_j \omega^{ij}~\texttt{mod q},~i\in[0, n-1]
    \end{equation}
    
    \vspace{-2.5mm}
    \begin{equation}\label{eq:INTT}
        \text{INTT} = a_i = n^{-1} \sum_{j=0}^{n-1} \tilde{a}_j \omega^{-ij}~\texttt{mod q},~i\in[0, n-1]
    \end{equation}

    In Eq.~\ref{eq:FNTT} and Eq.~\ref{eq:INTT}, $a(x)$ shows the polynomial in the algebraic domain, $\tilde{a}(x)$ is its NTT transformed counterpart, and $\omega$ is the primitive $n$-th root of unity. Kyber uses a $2n$-th primitive root, where $\omega=\zeta^2$ in Eq.~\ref{eq:FNTT} and Eq.~\ref{eq:INTT}. \malik{Hardware implementation of these equations involves zero padding,} which extends the polynomial length by appending zeros of equal size. While zero padding simplifies the implementation, it incurs higher computational overhead and memory usage~\cite{ntt_with_zero_padding}. An alternative approach, Negacyclic Wrapped Convolution (NWC), eliminates zero padding and reduces computation time but introduces additional pre- and post-processing complexities~\cite{ntt_with_NWC}. The Cooley-Tukey (CT) and Gentleman-Sande (GS) butterfly configurations inherently eliminate these extra processing costs. Consequently, CT-BU operations are defined in Eq.~\ref{eq:CT_exp1} and Eq.~\ref{eq:CT_exp2}, while the corresponding GS-BU operations are given in Eq.~\ref{eq:GS_exp1} and Eq.~\ref{eq:GS_exp2}. \malik{In these equations, $\texttt{u}$ and $\texttt{t}$ are input coefficients, $\texttt{U}$ and $\texttt{T}$ are output coefficients, $\mathrm{\zeta}$ is the twiddle factor, and dot ($\mathrm{\cdot}$) signifies integer multiplication.} For more details and derivations, readers are referred to~\cite{ntt_tutorial, ntt_complex_derivations}.

    %%=============================================%%
    %% CT-BU Equation
    %%=============================================%%

    \begin{empheq}[left={\makebox[4em][r]{{CT-BU} $\Rightarrow$ }\empheqlbrace}]{align}
    & \texttt{U} = \texttt{u} + (\texttt{t} \cdot \zeta)~\texttt{$\texttt{mod q}$} \label{eq:CT_exp1} \\
    & \texttt{T} = \texttt{u} - (\texttt{t} \cdot \zeta)~\texttt{$\texttt{mod q}$} \label{eq:CT_exp2}
    \end{empheq}

    %%=============================================%%
    %% GS-BU Equation
    %%=============================================%%

    \begin{empheq}[left={\makebox[4em][r]{{GS-BU} $\Rightarrow$ }\empheqlbrace}]{align}
        & \texttt{U} = (\texttt{u} + \texttt{t})~\texttt{$\texttt{mod q}$}~\label{eq:GS_exp1} \\
        & \texttt{T} = \zeta \cdot (\texttt{u} - \texttt{t})~\texttt{$\texttt{mod q}$}~\label{eq:GS_exp2}
    \end{empheq}    

    \malik{To integrate Eq.~\ref{eq:FNTT} and Eq.~\ref{eq:INTT} with the CT-BU and GS-BU equations for FNTT and INTT implementations, several algorithms are available in the literature. Algorithm~\ref{alg:fntt} presents a unified and iterative NTT approach. It integrates radix-2-based butterfly units of CT-BU and GS-BU (for Kyber parameters of $n=256$ \& $\texttt{q}=3329$), controlled by a one-bit \texttt{mode} signal. For FNTT, it performs seven CT-BU stages with $\texttt{len}=128,64,\dots,2$, and for INTT, it performs seven GS-BU stages with $\texttt{len}=2,4,\dots,128$. In each stage $s\in{1,\ldots,7}$, the input vector $a$ is processed in blocks of size $2\cdot\texttt{len}$, and each block uses a single twiddle factor $\zeta$, drawn from the precomputed tables (\texttt{zetas\_fwd} or \texttt{zetas\_inv}). After the seven stages of an INTT computation, it performs an additional rescaling step (lines~21--25), computing $a[i]\cdot n^{-1}~\texttt{mod q}$. 
    }

    %%=============================================%%
    %% Algorithm for NTT
    %%=============================================%%

    \begin{algorithm}[tb]
    \caption{\malik{Unified NTT for Kyber ($n=256$, $q=3329$)}}
    \label{alg:fntt}
    \begin{algorithmic}[1]

    \REQUIRE Length-$n$ vector $a=[a_0, \dots, a_{n-1}]$ ($a_i \in [0, q-1]$)

    \REQUIRE A one-bit signal $\texttt{mode}$ to compute $\text{FNTT}$ \& $\text{INTT}$
    
    \REQUIRE Precomputed twiddles: $\texttt{zetas\_fwd}[0..127]$ \& $\texttt{zetas\_inv}[0..127]$ (bit-reversed order)

    \ENSURE $a \leftarrow \text{FNTT}(a)$ if $\texttt{mode}=\text{FNTT}$ else $a \leftarrow \text{INTT}(a)$

    \STATE $k \gets 0$
    \STATE \textbf{if} $\texttt{mode}=\text{FNTT}$ \textbf{then} $\texttt{len}\gets128$, $Z\gets\texttt{zetas\_fwd}$ \textbf{else} $\texttt{len}\gets2$, $Z\gets\texttt{zetas\_inv}$

    \FOR{$s=1;\ s\le 7;\ s \gets s+1$} 
        
        \FOR{$\texttt{start}=0;\ \texttt{start}<256;\ \texttt{start}\gets \texttt{start}+2 \cdot \texttt{len}$}

            \STATE $\zeta \gets Z[k];\ k \gets k+1$

            \FOR{$j=\texttt{start};\ j<\texttt{start}+\texttt{len};\ j \gets j+1$}
                \STATE $u \gets a[j]$
                \STATE $t \gets a[j+\texttt{len}]$
                \IF{$\texttt{mode}=\text{FNTT}$} % CT: Eqs. (3),(4)
                    \STATE $t_\zeta \gets (t \cdot \zeta)~\texttt{mod q}$
                    \STATE $a[j] \gets (u + t_\zeta)~\texttt{mod q}$ \hfill\COMMENT{\texttt{Eq.~(3)}}
                    \STATE $a[j+\texttt{len}] \gets (u - t_\zeta)~\texttt{mod q}$ \hfill\COMMENT{\texttt{Eq.~(4)}}
                \ELSE % GS: Eqs. (5),(6)
                    \STATE $a[j] \gets (u + t)~\texttt{mod q}$ \hfill\COMMENT{\texttt{Eq.~(5)}}
                    \STATE $a[j+\texttt{len}] \gets (\zeta \cdot (u - t))~\texttt{mod q}$ \hfill\COMMENT{\texttt{Eq.~(6)}}
                \ENDIF
            \ENDFOR
        \ENDFOR
        \STATE \textbf{if} $\texttt{mode}=\text{FNTT}$ \textbf{then} $\texttt{len}\gets \texttt{len}/2$ \textbf{else} $\texttt{len}\gets 2 \cdot \texttt{len}$
    \ENDFOR
    
    \IF{$\texttt{mode}=\text{INTT}$}
    \FOR{$i=0;\ i<256;\ i\gets i+1$}
        \STATE $a[i]\gets (a[i]\cdot n^{-1})~\texttt{mod q}$ \hfill\COMMENT{\texttt{Rescaling step}}
    \ENDFOR
    \ENDIF
    \end{algorithmic}
    \end{algorithm}

    %%=============================================%%
    %% Section 3: DSE 
    %%=============================================%%

    \section{Design Space Exploration of CT-BU and GS-BU Butterfly Units}\label{sec:dse}

\begin{figure}[!]
            \flushleft
            \captionsetup[subfloat]{font=scriptsize}  % Reduces font size of subfigure captions
            \subfloat[\textsc{\textsc{BL-BU}}]{%
            \includegraphics[width=0.45\columnwidth]{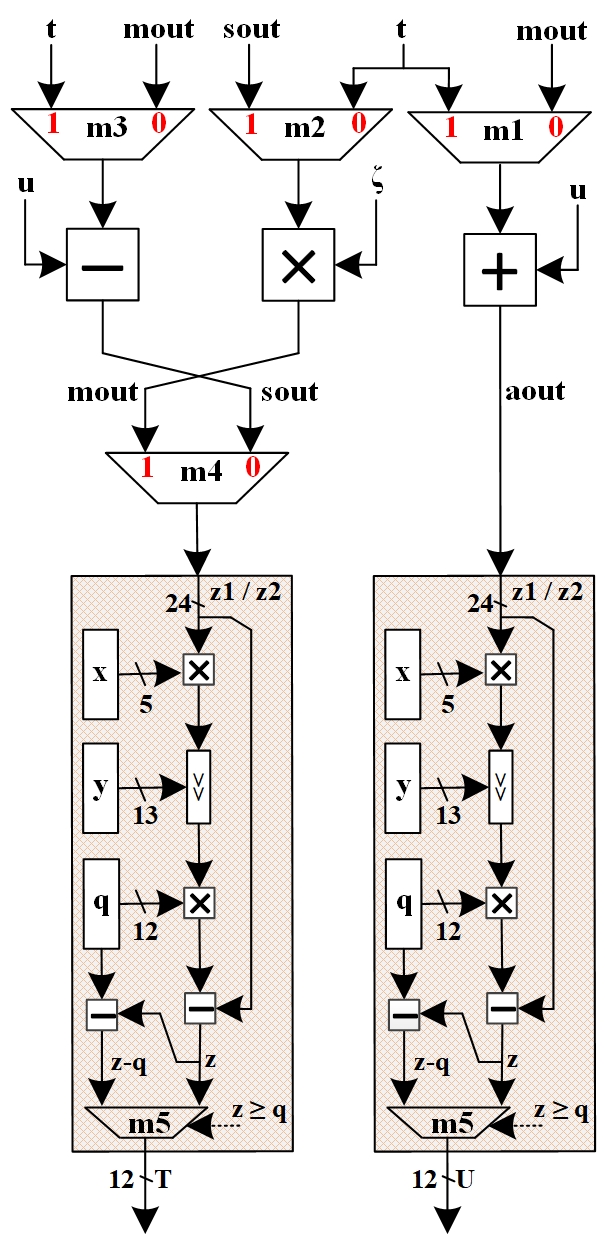}
            \label{fig:Baseline}
        }
            \subfloat[\textsc{8SP-BU}]{%
            \includegraphics[width=0.535\columnwidth]{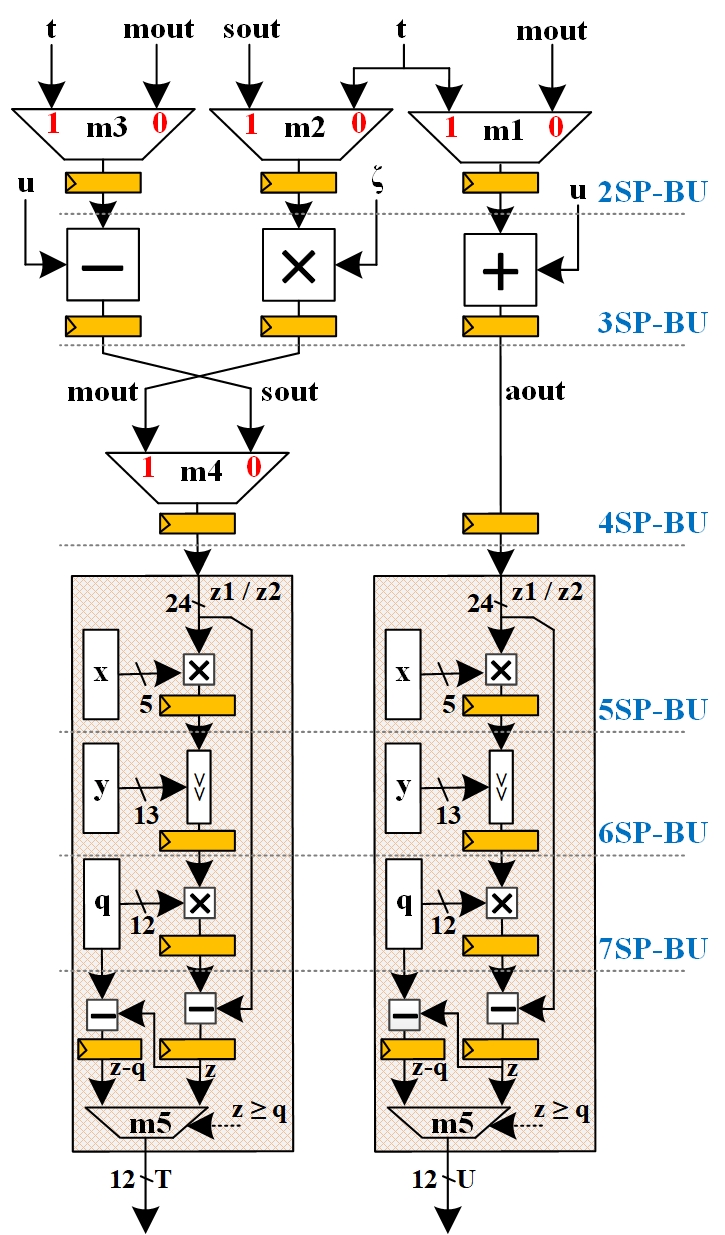}
            \label{fig:8SP}
        }
    \caption{\malik{Summary of the DSE process. The orange blocks denote the pipeline registers. The light-brown filled boxes show the modular reduction ($\texttt{mod q}$ with $\texttt{q}=3329$). }
    % The pipelined architectures (from \textsc{2SP-BU} to \textsc{8SP-BU}) are identical to the baseline (\textsc{\textsc{BL-BU}}) of Fig.~\ref{fig:Baseline}, only the difference is the placement of pipeline registers. Therefore, the placement of pipeline registers for the architectures of \textsc{2SP-BU} to \textsc{7SP-BU} are highlighted in Fig.~\ref{fig:8SP}. 
    }
    \label{fig:overall}
    \end{figure}

    The DSE in our work focuses on optimizing the circuit frequency of the unified design of the CT-BU and GS-BU butterfly units. This optimization is achieved by extensively pipelining the data path, including the multiplexers, arithmetic blocks, and delving into the modular reduction blocks of the unified design of CT-BU and GS-BU butterfly units. The summary of the DSE process is shown in Fig.~\ref{fig:overall}. It involves eight architectures: (i) a baseline non-pipelined architecture and (ii) seven pipelined architectures. The details of the implemented architectures are provided below.

    \vspace{+1.5mm}
     \begin{itemize}
        \item {Baseline $\Rightarrow$}\hskip 9pt \{\enspace \textsc{\textsc{BL-BU}} \\
        
        \item {Pipelined $\Rightarrow$} 
        \(
          \left\{
            \begin{tabular}{@{\enspace}l}
              \textsc{2SP-BU} (2-stage pipelined butterfly) \\
              \textsc{3SP-BU} (3-stage pipelined butterfly)\\
              \textsc{4SP-BU} (4-stage pipelined butterfly)\\
              \textsc{5SP-BU} (5-stage pipelined butterfly)\\
              \textsc{6SP-BU} (6-stage pipelined butterfly)\\
              \textsc{7SP-BU} (7-stage pipelined butterfly)\\
              \textsc{8SP-BU} (8-stage pipelined butterfly)
            \end{tabular}
            \right.
        \)
    \end{itemize}

    %%=============================================%%
    %% I/O Interface of Baseline and Pipelined Butterfly Units
    %%=============================================%%

    % \subsection{I/O Interface of Baseline and Pipelined Butterfly Units}     The baseline and pipelined butterfly units share an identical I/O interface, which is not shown in Fig.~\ref{fig:overall} for simplicity. This interface comprises three 12-bit inputs and two 12-bit outputs. The inputs are $\texttt{u}$, $\texttt{t}$, and $\mathrm{\zeta}$, while the outputs are $\texttt{U}$ and $\texttt{T}$. The 12-bit size for each I/O signal is selected based on the Kyber coefficient length. Specifically, $\texttt{u}$ and $\texttt{t}$ are input coefficients, while $\mathrm{\zeta}$ denotes the twiddle factors for the FNTT and INTT operations. The twiddle factors are complex exponential coefficients necessary for manipulating data between the time and NTT domains. Similarly, $\texttt{U}$ and $\texttt{T}$ correspond to the outputs of Eq. \ref{eq:CT_exp1}, \ref{eq:CT_exp2}, and Eq. \ref{eq:GS_exp1}, \ref{eq:GS_exp2}, respectively. For simplification, the clock, reset, and control signals (for the routing multiplexers) are not shown in Fig.~\ref{fig:overall}.

    \vspace{+2.5mm}
    The baseline and pipelined butterfly units share an identical I/O interface, omitted from Fig.~\ref{fig:overall} for simplicity. This interface includes three 12-bit inputs ($\texttt{u}$, $\texttt{t}$, and $\mathrm{\zeta}$) and two 12-bit outputs ($\texttt{U}$ and $\texttt{T}$) with the bit-width chosen to match the Kyber coefficient length. Here, $\texttt{u}$ and $\texttt{t}$ are input coefficients, while $\mathrm{\zeta}$ represents the twiddle factors used in FNTT and INTT operations. These twiddle factors are complex exponential coefficients essential for transforming data between the time and NTT domains. The outputs $\texttt{U}$ and $\texttt{T}$ correspond to the results of Eq.~\ref{eq:CT_exp1}, \ref{eq:CT_exp2}, and Eq.~\ref{eq:GS_exp1}, \ref{eq:GS_exp2}. For clarity, clock, reset, and control signals (e.g., for routing multiplexers) are also excluded from Fig.~\ref{fig:overall}.

    %%=============================================%%
    %% Baseline Butterfly
    %%=============================================%%
    
    \subsection{Unified Baseline Architecture (\textsc{\textsc{BL-BU}})}

   The unified baseline architecture (\textsc{BL-BU}) for the CT-BU and GS-BU units is illustrated in Fig.~\ref{fig:Baseline}. It comprises three arithmetic operators (an adder, a subtractor, and a multiplier) along with two modular reduction units. These operations are implemented using Verilog’s built-in operators, employing the 12-bit Kyber coefficient size to minimize hardware complexity. \malik{We adopt the Barrett method for modular reduction (\texttt{mod q}), consistent with~\cite{safi_ntt}, but introduce a key enhancement: our \texttt{mod q} unit is deeply pipelined to maximize operating frequency, unlike the non-pipelined version in~\cite{safi_ntt}. Barrett is chosen over more efficient methods like Plantard or Montgomery due to its seamless integration with the butterfly unit. Specifically, Barrett maintains the original coefficient width, avoiding the word-size expansion required by Plantard. Although Montgomery is efficient, it operates in a separate arithmetic domain and incurs overhead from pre- and post-conversions.   Thus, Barrett offers a balanced trade-off between performance and architectural simplicity, making it a zero-overhead fit for our design.}

    %   In Fig.~\ref{fig:overall}, the \texttt{mod q} blocks are highlighted with brown pattern-filled boxes, with constants denoted by $\texttt{x}$, $\texttt{y}$ and $\texttt{q}$. In addition to the I/O interface and the arithmetic operators, the \textsc{BL-BU} design includes four $2\times 1$ multiplexers ($\texttt{m1}$, $\texttt{m2}$, $\texttt{m3}$, $\texttt{m4}$) controlled by a single signal. When the control signal is \texttt{0}, the circuit computes FNTT (Eq.~\ref{eq:CT_exp1} and Eq.~\ref{eq:CT_exp2}). When the control signal is \texttt{1}, it computes INTT (Eq.~\ref{eq:GS_exp1} and Eq.~\ref{eq:GS_exp2}). As far as the computational cost of Fig.~\ref{fig:Baseline} is concerned, it takes one clock cycle to process each 12-bit Kyber polynomial coefficient per stage, as the \textsc{BL-BU} data path lacks pipeline registers. \malik{The total number of stages required to implement Kyber NTT is $\mathrm{\log_2(n)}-1$, and each stage requires $\frac{n}{2}$ coefficients, where $n=256$.} Therefore, at least 896 ($7\times \frac{256}{2}$) clock cycles are needed to compute the FNTT of Kyber using Algorithm~\ref{alg:fntt}. Moreover, our \textsc{BL-BU} requires four additional clock cycles. We want to highlight that, without pipelining, the total data path delay is 23.95$ns$, whereas the routing and logic delays are 13.74 and 10.21$ns$, respectively. Therefore, pipelining is needed to minimize these delays. More details will be provided later in~section~\ref{sec:results_and_comparisons}.   

    In Fig.~\ref{fig:overall}, the \texttt{mod q} blocks are shown as brown pattern-filled boxes, with constants $\texttt{x}$, $\texttt{y}$, and $\texttt{q}$. 
    The \textsc{BL-BU} design includes three arithmetic units and four $2\times 1$ multiplexers ($\texttt{m1}$–$\texttt{m4}$), all controlled by a single signal. 
    When the signal is set to \texttt{0}, the circuit performs FNTT (Eq.~\ref{eq:CT_exp1}, \ref{eq:CT_exp2}); when set to \texttt{1}, it performs INTT (Eq.~\ref{eq:GS_exp1}, \ref{eq:GS_exp2}). Each 12-bit Kyber coefficient is processed in one clock cycle per stage, as \textsc{BL-BU} lacks pipeline registers. \malik{For Kyber ($n=256$), the NTT requires $\log_2(n)-1$ stages and $\frac{n}{2}$ coefficients per stage, totaling 896 cycles.} An additional four cycles are needed for control. Without pipelining, the data path delay is 23.95$ns$, with routing and logic delays of 13.74$ns$ and 10.21$ns$, respectively. Therefore, pipelining is needed to minimize these delays. More details will be provided later in~Section~\ref{sec:results_and_comparisons}.

    %%==================================%%
    % Pipelined Butterfly Architectures
    %%==================================%%
    
   \subsection{Unified Pipelined Architectures (\textsc{2SP-BU} to \textsc{8SP-BU})}

   Each pipelined butterfly unit includes the same arithmetic and modular reduction components as the \textsc{BL-BU}. 
   The key distinction lies in the strategic placement of registers within the data path, which enables pipelining and improves throughput.

    \begin{itemize}
   \item \textit{Coarse-grained pipelining:} This approach begins by inserting registers into the \textsc{BL-BU} data path, excluding the $\texttt{mod q}$ block. It yields three designs: (i) \textsc{2SP-BU}, (ii) \textsc{3SP-BU}, and (iii) \textsc{4SP-BU}, all featuring non-pipelined $\texttt{mod q}$ circuits. Among them, \textsc{4SP-BU} delivers the best performance, achieving a 3.93$\times$ frequency gain with a 1.15$\times$ area overhead by placing registers after the routing multiplexers and arithmetic units.

    \item  \textit{Fine-grained pipelining:} It further optimizes \textsc{4SP-BU} by inserting registers into the $\texttt{mod q}$ block, resulting in four new designs: \textsc{5SP-BU} to \textsc{8SP-BU} (Fig.~\ref{fig:8SP}). Among all eight, \textsc{8SP-BU} is the most optimized, achieving a 6.47$\times$ frequency improvement with 1.32$\times$ area overhead compared to \textsc{BL-BU} (baseline architecture). Details of the experimental results will be discussed in Section~\ref{sec:results_and_comparisons}. 
    \end{itemize}
        
      Finally, it is essential to note that we have not employed any automated approach to insert pipeline registers in the data path of the realized architectures. Instead, the placement of pipeline registers is determined based on evaluating the critical path, which will be discussed later in Section~\ref{sec:results_and_comparisons}.

    %%==================================%%
    % Section 4: Proposed Memory Parallelization Technique
    %%==================================%%
    
    \section{Proposed Memory Parallelization Technique} \label{sec:proposed_memory_parallel_approach}

    \malik{Our memory-parallelization approach enables partial parallelism in fully iterative NTT designs through the following strategies:}

    \begin{itemize}
        \item Pairing two Kyber coefficients at each memory address to reduce access overhead,
        \item Using four $\frac{n}{4}$-sized memory blocks in parallel to minimize clock cycles, and
        \item Employing two butterfly units to balance hardware cost and computational efficiency.
    \end{itemize}

\begin{figure*}[!t]       
    
            \centering
            \captionsetup[subfloat]{font=scriptsize}  % Reduces font size of subfigure captions
            
            \subfloat[Pairing Coefficients (for FNTT)]{%
            \includegraphics[height = 7cm, width=0.5\columnwidth]{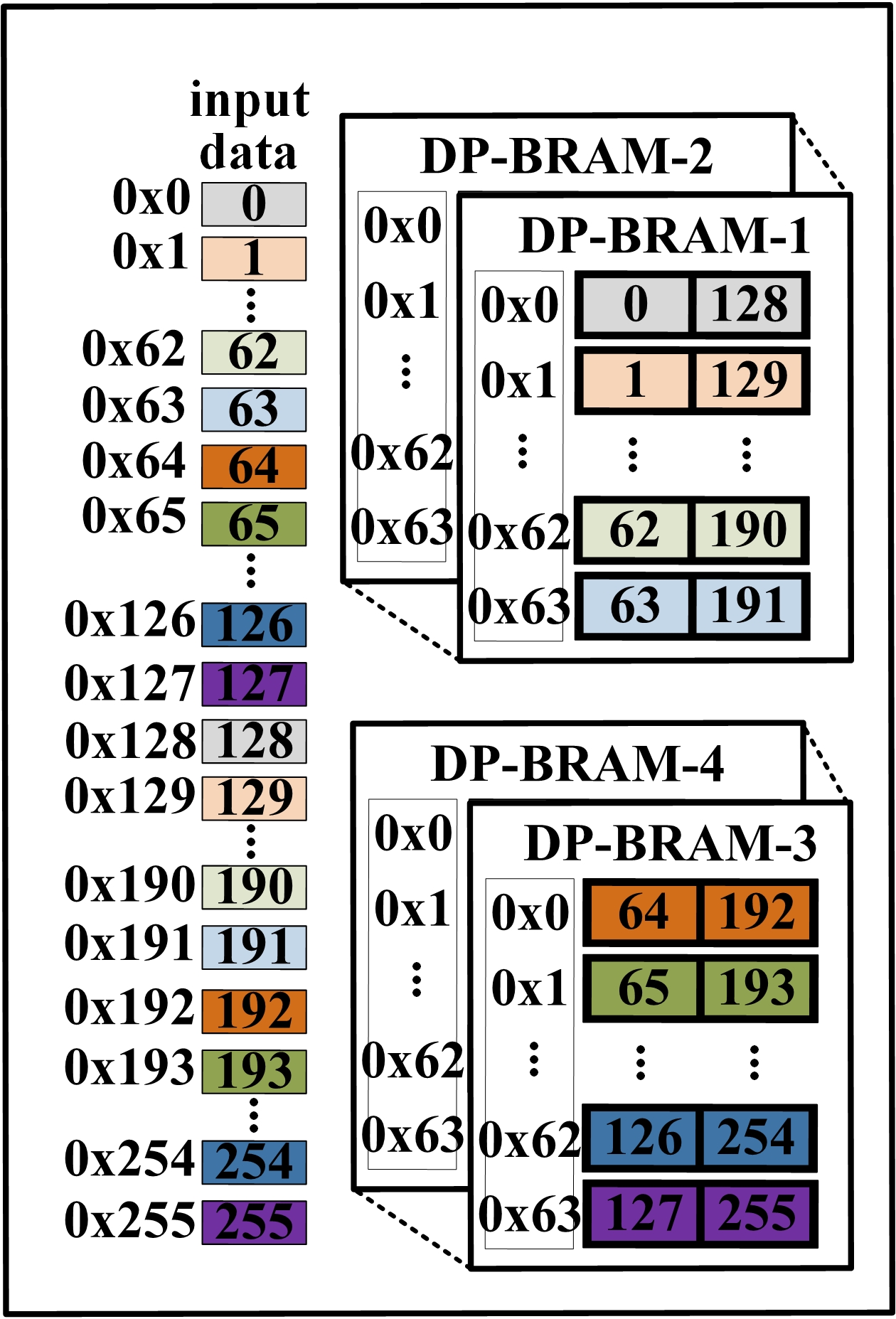}
            \label{fig:pairing_coefficients_fntt}
        }
    % \hfill
    % \hspace{0.05\textwidth}
            \centering
            \subfloat[Pairing Coefficients (for INTT)]{%
            \includegraphics[height = 7cm, width=0.5\columnwidth]{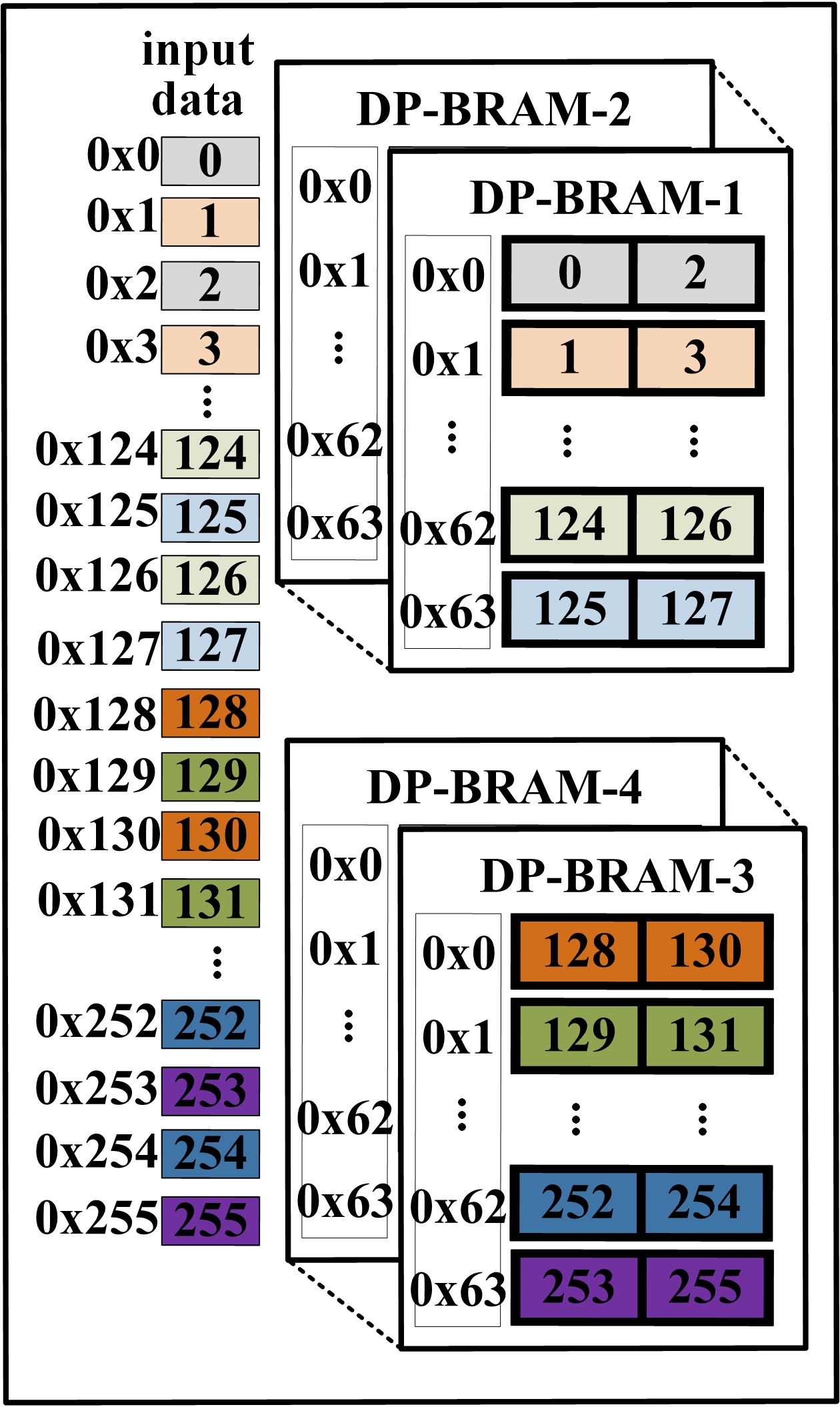}
            \label{fig:pairing_coefficients_intt}
        }
    % \hfill
    % \hspace{0.05\textwidth}
            \centering
            \subfloat[Adjacent Coefficient Processing]{%
            \includegraphics[height = 7cm, width=0.5\columnwidth]{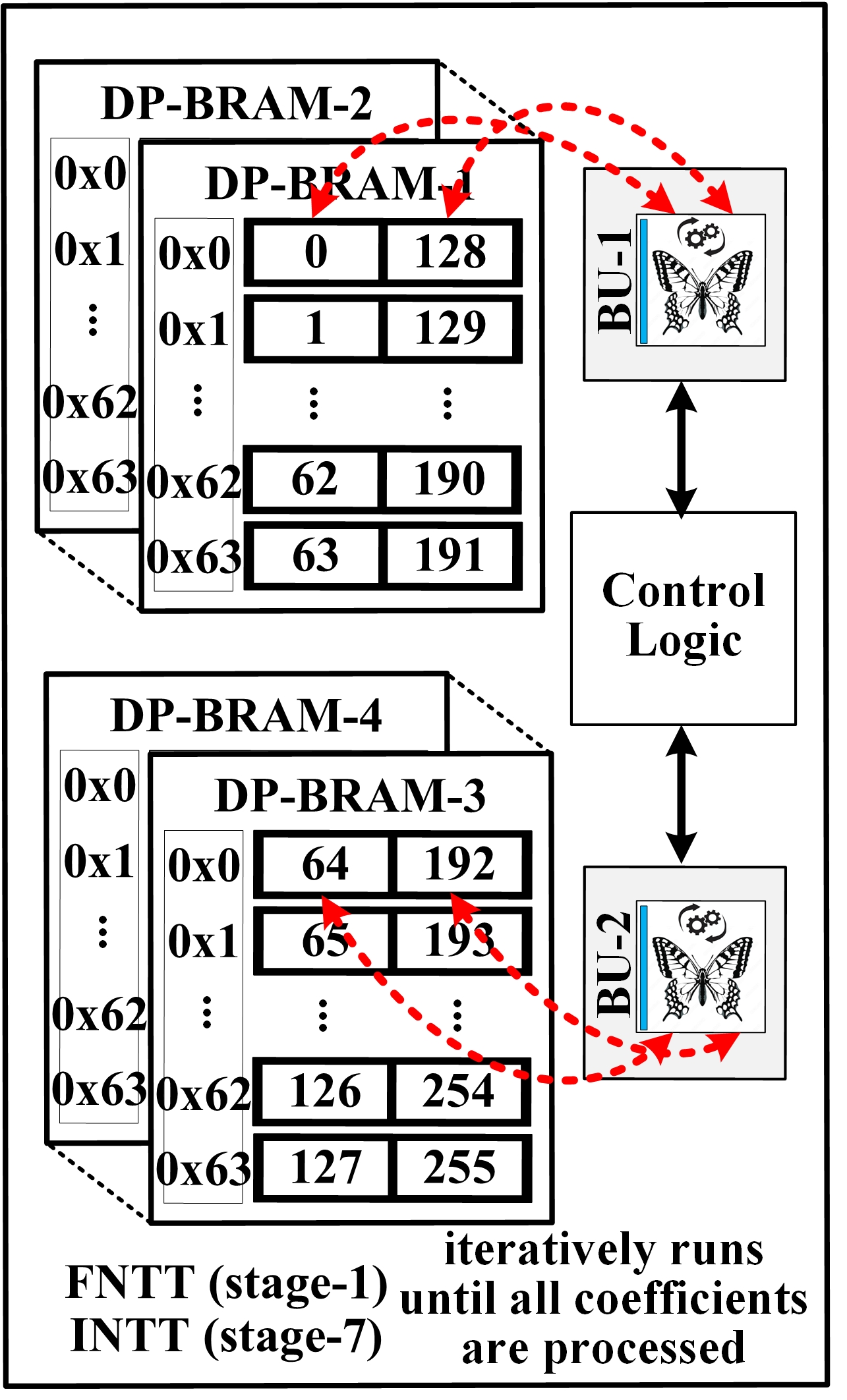}
            \label{fig:adjacent_coefficient_processing}
        }
    % \hfill
    % \hspace{0.05\textwidth}
            \centering
            \subfloat[Coefficients Reordering]{%
            \includegraphics[height = 7cm, width=0.5\columnwidth]{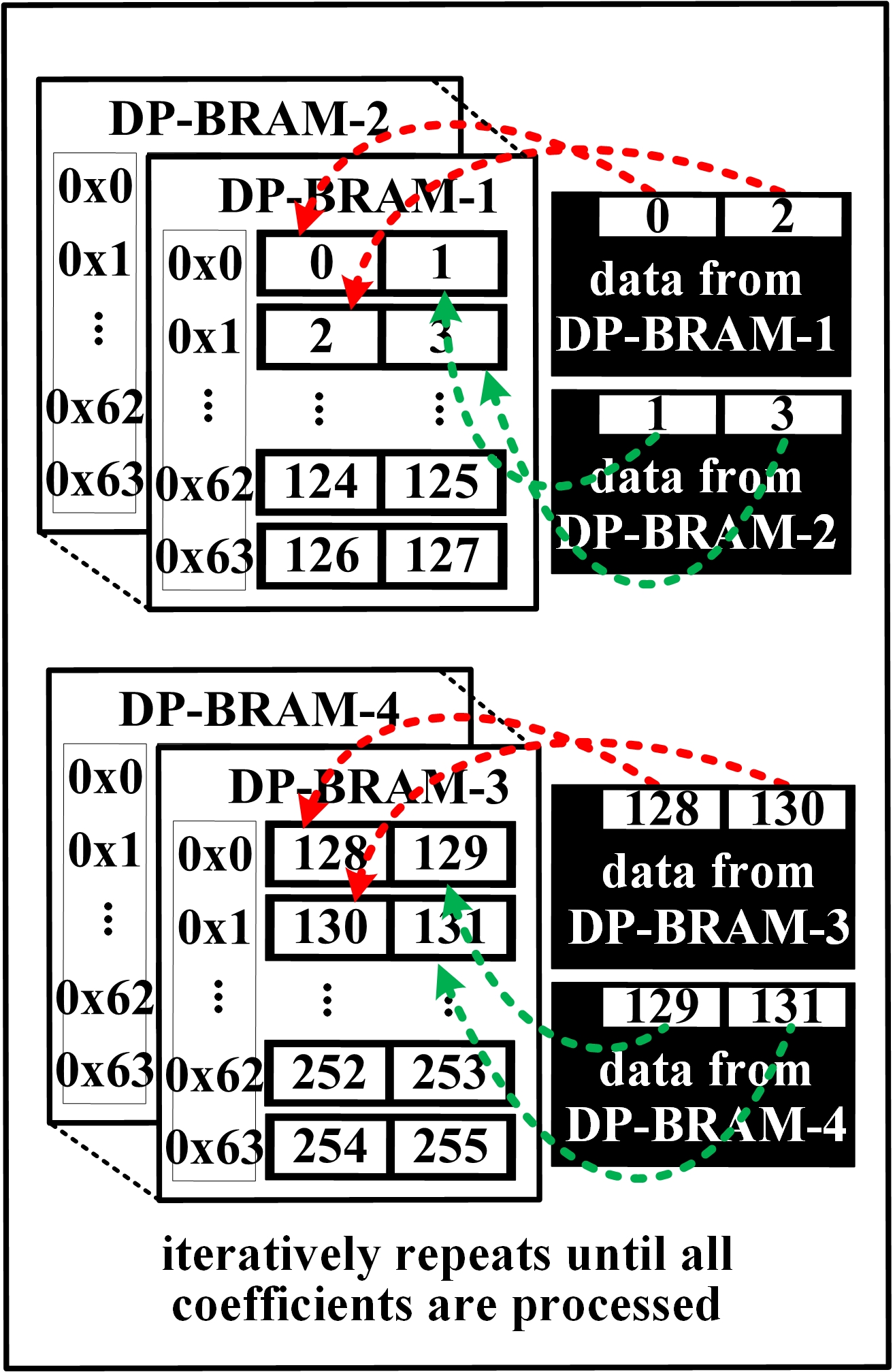}
            \label{fig:coefficients_reoredring}
        }
    
    % \hfill
    % \hspace{0.05\textwidth}
            \centering
            \subfloat[Swapping (\malik{offset~$(\delta)\ge \frac{n}{4}$})]{%
            \includegraphics[height = 7cm, width=\columnwidth]{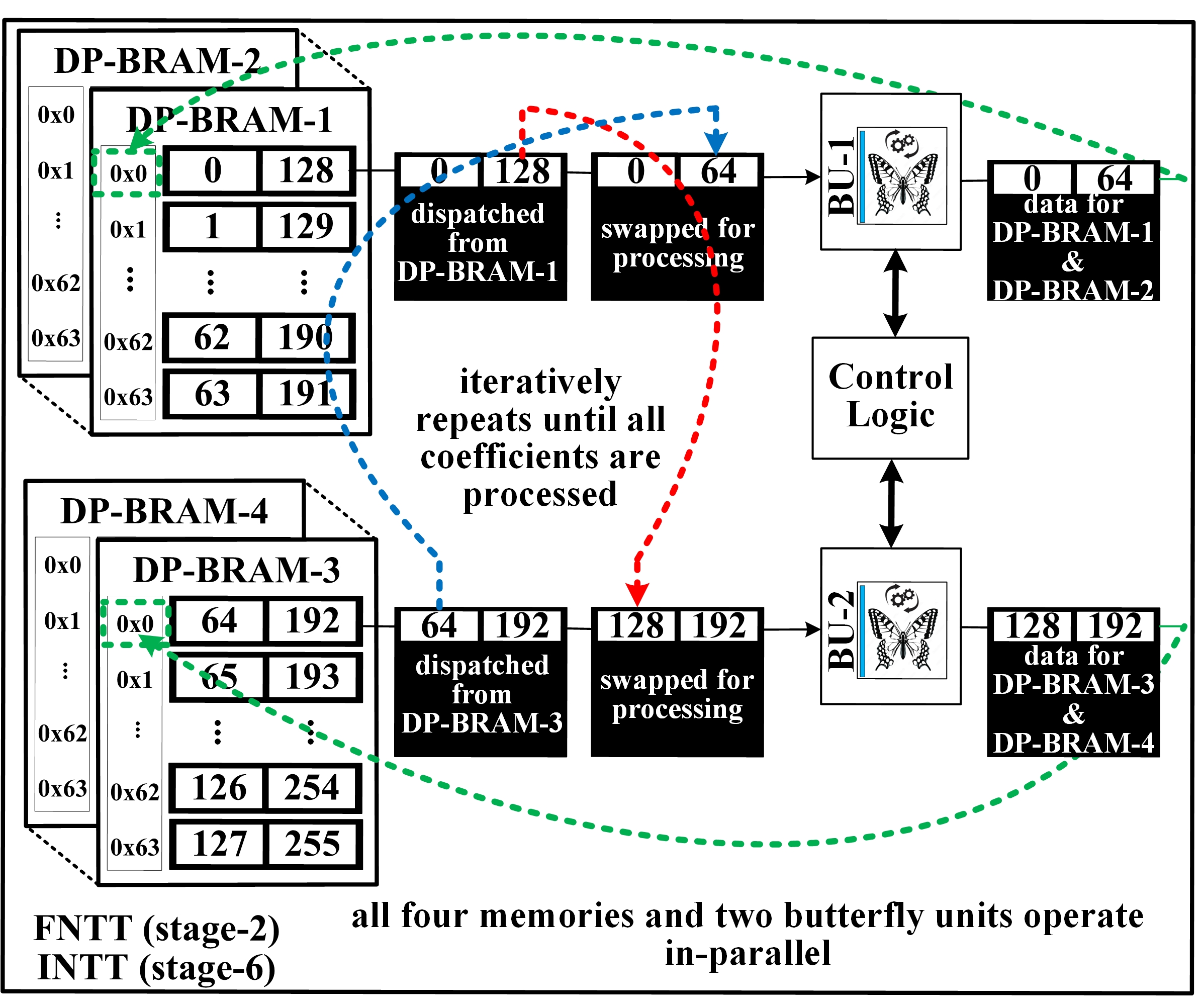}
            \label{fig:swapping_offset_64}
        }
    % \hfill
    % \hspace{0.05\textwidth}
            \centering
            \subfloat[Swapping (\malik{offset~$(\delta)< \frac{n}{4}$})]{%
            \includegraphics[height = 6.88cm, width=\columnwidth]{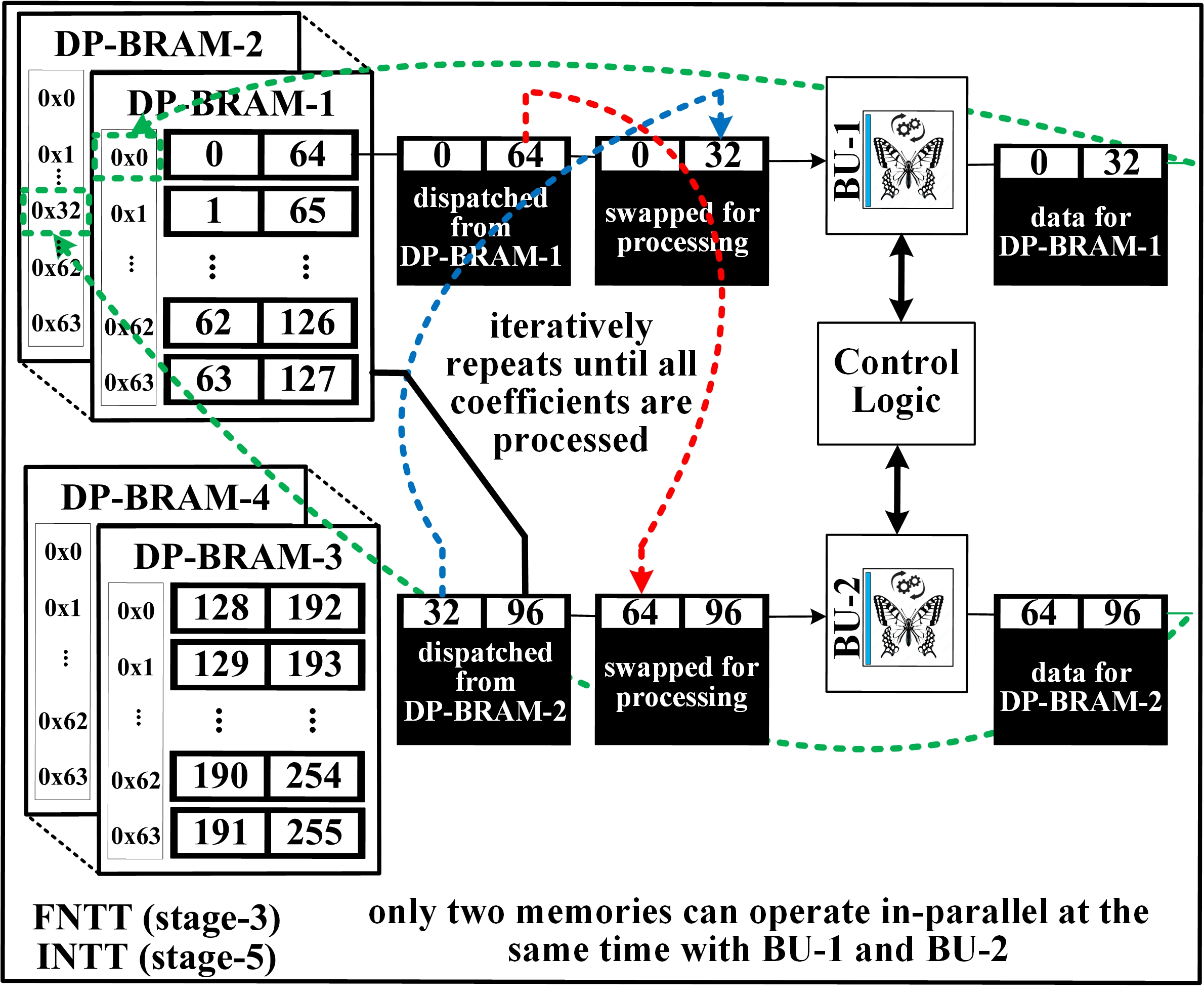}
            \label{fig:swapping_offset_32}
        }
    \caption{Overview of the proposed memory parallelization approach.}
    % \vspace{-5mm}
    \label{fig:our_mem_parallel_approach}
    \end{figure*}

    Fig.~\ref{fig:our_mem_parallel_approach} illustrates the \malik{architectural} components of this approach: (i) coefficient pairing (Fig.~\ref{fig:pairing_coefficients_fntt} and~\ref{fig:pairing_coefficients_intt}), and (ii) optimized parallel memory access (Fig.~\ref{fig:adjacent_coefficient_processing},~\ref{fig:coefficients_reoredring},~\ref{fig:swapping_offset_64}, and~\ref{fig:swapping_offset_32}). 
    The four memory blocks, each of size $\frac{n}{4}$, are labeled $\texttt{DP-BRAM-1}$ through $\texttt{DP-BRAM-4}$, while the two butterfly units are denoted BU-1 and BU-2. Further details are provided in Sections~\ref{subsec:pairing_coefficients} and~\ref{subsec:parallel_memory_access}.

    %%==================================%%
    % Figure for General Idea
    %%==================================%%

    % Fig.~\ref{fig:our_mem_parallel_approach} describes the building blocks of our memory parallelization approach: (i) pairing coefficients (in Fig.~\ref{fig:pairing_coefficients_fntt} and~\ref{fig:pairing_coefficients_intt}), and (ii) optimizing the parallel memory access (see Fig.~\ref{fig:adjacent_coefficient_processing},~\ref{fig:coefficients_reoredring},~\ref{fig:swapping_offset_64},~and~\ref{fig:swapping_offset_32}). In these figures, the four memory blocks, each of size $\frac{n}{4}$, are labeled $\texttt{DP-BRAM-1}$, $\texttt{DP-BRAM-2}$, $\texttt{DP-BRAM-3}$, and $\texttt{DP-BRAM-4}$. Similarly, the two butterfly units (if needed) are shown in BU-1 and BU-2. The corresponding details are in section~\ref{subsec:pairing_coefficients} and section~\ref{subsec:parallel_memory_access}.  
    
    %%==================================%%
    % Pairing Coefficients
    %%==================================%%
    
    \subsection{Pairing Coefficients} \label{subsec:pairing_coefficients}

   % For FNTT computations, input coefficients must be pre-arranged such that, for $n = 256$, the first $\frac{n}{2}-1$ coefficients are paired with those from $\frac{n}{2}$ to $n-1$. This arrangement is illustrated in Fig.~\ref{fig:pairing_coefficients_fntt}, where input data with similar color blocks are grouped and mapped onto memory units. Similarly, the pairing coefficients for INTT are depicted in Fig.~\ref{fig:pairing_coefficients_intt}, with the key difference being the offset, which determines the memory address step size to access the coefficients at various computation stages. For FNTT, in Fig.~\ref{fig:pairing_coefficients_fntt}, the pairing is considered based on the initial offset value of 128, while for INTT, in Fig.~\ref{fig:pairing_coefficients_intt}, it begins at 2. More specifically, in FNTT, the offset starts at 128 and halves at each stage until it reaches 2 (for Kyber). In contrast, in INTT, the offset starts at 2 and doubles at each stage, following the sequence $\mathrm{2, 4, 8, \dots, 128}$.

    For FNTT computations, input coefficients must be pre-arranged such that, for $n = 256$, the first $\frac{n}{2}-1$ coefficients are paired with those from $\frac{n}{2}$ to $n-1$. This arrangement is illustrated in Fig.~\ref{fig:pairing_coefficients_fntt}, where input data (similarly colored blocks) are grouped and mapped onto memory units. The INTT pairing, shown in Fig.~\ref{fig:pairing_coefficients_intt}, differs primarily in the offset, which determines the memory address step size at each computation stage.  More specifically, in FNTT, the offset starts at 128 and halves at each stage until it reaches 2 (for Kyber). In contrast, in INTT, the offset starts at 2 and doubles at each stage, following the sequence $\mathrm{2, 4, 8, \dots, 128}$.

    % For FNTT computations, input coefficients need to be pre-arranged so that, for $n = 256$, the first $\frac{n}{2}-1$ coefficients are paired with those from $\frac{n}{2}$ to $n-1$. This is shown in~Fig.~\ref{fig:pairing_coefficients_fntt}, where the input data with similar color blocks are grouped and consequently mapped onto the memory units. Similarly, the pairing coefficients for INTT are presented in Fig.~\ref{fig:pairing_coefficients_intt}; the only difference is the offset, which refers to the memory address step size used to access coefficients at different stages of the computation. For FNTT, the offset is 128, while for INTT, the offset value is 2.

    %%==================================%%
    % Parallel Memories Access 
    %%==================================%%
    
    \subsection{Optimizing Parallel Memory Access} \label{subsec:parallel_memory_access}

    This section describes the use of four $\frac{n}{4}$-sized memory blocks operating in parallel with two butterfly units to minimize computation time. Before FNTT or INTT computation begins, $\texttt{DP-BRAM-1}$ and $\texttt{DP-BRAM-2}$ store identical paired coefficients, while $\texttt{DP-BRAM-3}$ and $\texttt{DP-BRAM-4}$ contain similar pairs. A total of $\frac{n}{4}$ clock cycles is required to load $n$ coefficients into the memory blocks in parallel. To fully exploit this parallelism, all memory blocks are used simultaneously for read and write operations. However, this introduces a critical challenge: coefficient swapping prior to butterfly processing to ensure correct computation. To address this, we employ two key techniques: (i) adjacent coefficient processing and (ii) coefficient swapping before processing. These approaches are discussed in detail in the following sections.

    %%==================================%%
    % Adjacent coefficient processing 
    %%==================================%%
    \subsubsection{Adjacent coefficient processing}\label{subsubsec:adjacent_coefficient_processing}

    As illustrated in Fig.~\ref{fig:adjacent_coefficient_processing}, input coefficients are stored in pairs within memory units (see Section~\ref{subsec:pairing_coefficients}), requiring adjacent coefficient processing only during the initial stage of FNTT or the final stage of INTT. The core idea is to process each coefficient pair stored at the same memory address using two butterfly units in parallel. Specifically, $\texttt{DP-BRAM-1}$ and $\texttt{DP-BRAM-2}$ feed BU-1, while $\texttt{DP-BRAM-3}$ and $\texttt{DP-BRAM-4}$ feed BU-2, as visually highlighted by red dotted lines. For example, at address $\mathrm{0x0}$, BU-1 processes the pair ($\mathrm{0, 128}$) and BU-2 handles ($\mathrm{64, 192}$); at $\mathrm{0x1}$, BU-1 computes ($\mathrm{1, 129}$) and BU-2 processes ($\mathrm{65, 193}$). This pattern continues across all memory addresses, completing the stage in $\frac{n}{4}$ clock cycles.

    %%==================================%%
    % Coefficient Swapping  
    %%==================================%%

    \subsubsection{Coefficient swapping}\label{subsubsec:swapping}

    \malik{In all stages of the FNTT and INTT (except for the initial stage of the FNTT and the final stage of the INTT), coefficient swapping is required prior to processing when small distributed memories (e.g., $\tfrac{n}{4}$-sized blocks in our approach) are used for parallelism. An effective method is to implement swapping using purely combinational logic, controlled by an offset or memory step-size parameter ($\delta$) generated by the control unit. We adopt this strategy, which will be further detailed in Section~\ref{subsec:swap_unit_CU}.} In our parallel memory-access approach, coefficient swapping is performed in the following two cases:

    \begin{itemize}
    \item[(i)] \textbf{\malik{$\mathbf{Offset~(\delta) \ge \frac{n}{4}}$:}} All four memory blocks and both butterfly units (BU-1 and BU-2) can be fully utilized when the offset $\delta \ge \frac{n}{4}$. For example, in Fig.~\ref{fig:swapping_offset_64}, at address $\mathrm{0x0}$, $\texttt{DP-BRAM-1}$ and $\texttt{DP-BRAM-2}$ store the coefficient pair ($\mathrm{0, 128}$), while $\texttt{DP-BRAM-3}$ and $\texttt{DP-BRAM-4}$ store ($\mathrm{64, 192}$). Since each stage depends on the output of the previous one, stage-2 of FNTT requires coefficients computed from stage-1. To ensure correct processing, BU-1 handles the swapped pair ($\mathrm{0, 64}$), and BU-2 processes ($\mathrm{128, 192}$). This swapping mechanism is highlighted with red and blue dotted lines in Fig.~\ref{fig:swapping_offset_64}, while the output coefficients are stored at the same memory address ($\mathrm{0x0}$), as shown by the green dotted line. This process iterates across all memory addresses, completing the stage in $\frac{n}{4}$ clock cycles.

    \item[(ii)] \textbf{\malik{$\mathbf{Offset~(\delta) < \frac{n}{4}}$:}} As discussed in Section~\ref{subsec:pairing_coefficients}, the FNTT offset starts at 128 and halves each stage until it reaches 2. In early stages (offsets 128 and 64), all four BRAMs and both butterfly units are fully utilized (see Figs.~\ref{fig:adjacent_coefficient_processing} and~\ref{fig:swapping_offset_64}). However, in later stages, the offset becomes smaller than the memory size ($\frac{n}{4}$), limiting parallel access. Additionally, only two butterfly units are available, further restricting parallelism. For example, in Fig.~\ref{fig:swapping_offset_32}, during FNTT stage-3 and INTT stage-5, coefficients from addresses $\mathrm{0x0}$ and $\mathrm{0x32}$ in $\texttt{DP-BRAM-1}$ and $\texttt{DP-BRAM-2}$ are ($\mathrm{0, 64}$) and ($\mathrm{32, 96}$), respectively. BU-1 processes ($\mathrm{0, 32}$), and BU-2 handles ($\mathrm{64, 96}$), leaving $\texttt{DP-BRAM-3}$ and $\texttt{DP-BRAM-4}$ idle. Once processing completes, the roles alternate, continuing this pattern until the final Kyber stage.
\end{itemize}

    \malik{Based on the two cases above, we summarize that for $\delta \ge \frac{n}{4}$, the design achieves full parallelism, with all four memory units operating concurrently. For $\delta < \frac{n}{4}$, it achieves partial parallelism, where two memory units are active and two remain idle. This memory-parallelization strategy benefits iterative NTT designs by enabling partial parallelism without increasing memory overhead. For Kyber ring size $n$, using two $12\times n$-sized memory units is equivalent to four $24\times \frac{n}{4}$-sized units. Here, $12\times n$ indicates packing one 12-bit Kyber coefficient per memory address across $n$ addresses, while $24\times \frac{n}{4}$ packs two 12-bit coefficients per address over $\frac{n}{4}$ addresses.}

    \malik{As a result of memory parallelization, the final stage of FNTT and INTT produces coefficients that appear scattered across multiple memories rather than stored sequentially. Although Fig.~\ref{fig:our_mem_parallel_approach} does not explicitly illustrate this scattering, an 8-point NTT case study in Section~\ref{subsec:case_study} clarifies the behavior. To restore the natural order, coefficient reordering is applied, as shown in Fig.~\ref{fig:coefficients_reoredring}, effectively mapping distributed outputs back to their sequential form.}

    %%==================================%%
    % 8-point NTT (case study)  
    %%==================================%%

    \subsection{8-point NTT (case study)} \label{subsec:case_study}

    %%==================================%%
    % Figure for 8_point_ntt_cstudy
    %%==================================%%

    \begin{figure}[tb]
    \centering
        \vspace{-3.5mm}
        \includegraphics[width=\columnwidth]{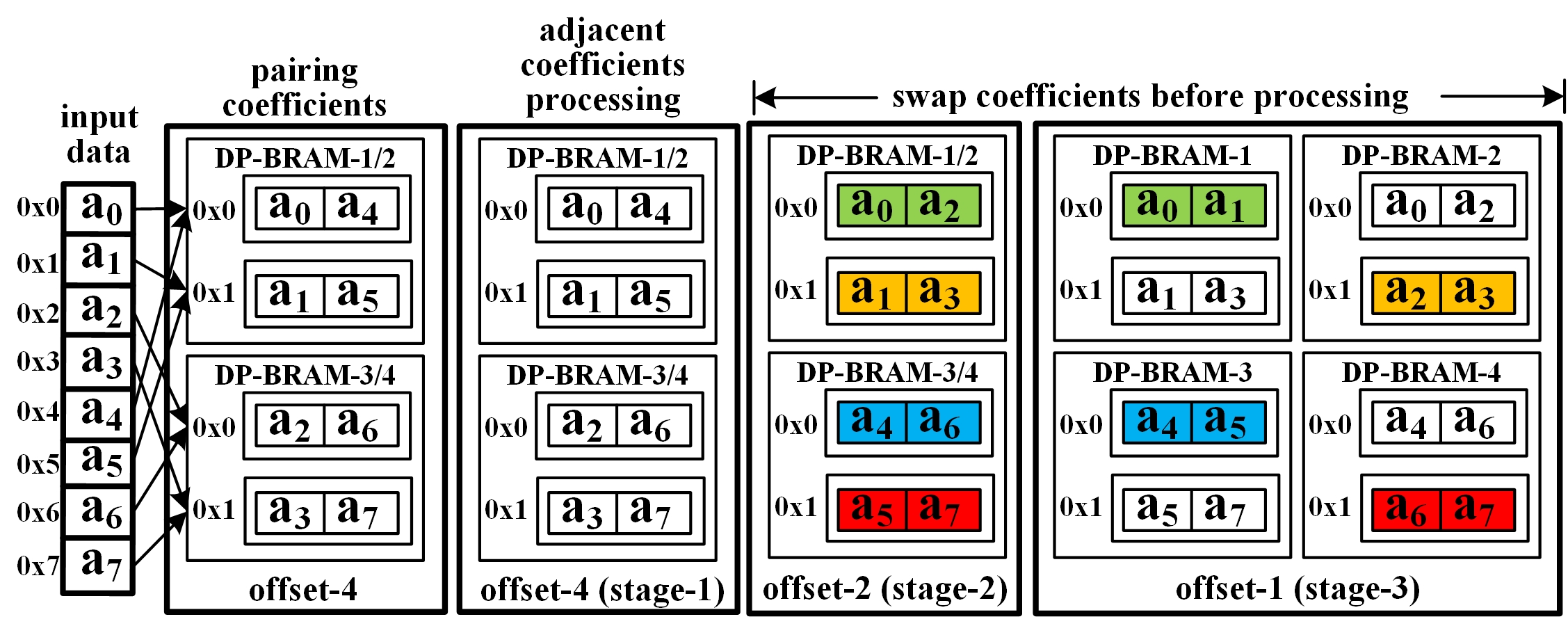}
        \caption{Case study: example of an 8-point NTT. 
        At stage-3, the colored filled boxes (green, orange, blue, and red) show the scattered distribution of the final FNTT coefficients in multiple memories.   
        }
        \vspace{-3mm}
    \label{fig:8_point_ntt_cstudy}
    \end{figure}

    We applied the concepts discussed previously in Section~\ref{subsec:pairing_coefficients} and Section~\ref{subsec:parallel_memory_access} to an 8-point NTT as a case study in Fig.~\ref{fig:8_point_ntt_cstudy}, where the total required stages are 3 and the number of butterfly operations to process in each stage is 4. 
    The leftmost part of Fig.~\ref{fig:8_point_ntt_cstudy} illustrates the initial input coefficients ($\mathrm{a_0, a_1, \dots, a_7}$) stored sequentially at addresses $\mathrm{0x0}$ to $\mathrm{0x7}$. These coefficients are paired using an offset of 4 and loaded into four DP-BRAMs: $\texttt{DP-BRAM-1}$ and $\texttt{DP-BRAM-2}$ store ($\mathrm{a_0, a_4}$) at address $\mathrm{0x0}$ and ($\mathrm{a_1, a_5}$) at $\mathrm{0x1}$, while $\texttt{DP-BRAM-3}$ and $\texttt{DP-BRAM-4}$ store ($\mathrm{a_2, a_6}$) at $\mathrm{0x0}$ and ($\mathrm{a_3, a_7}$) at $\mathrm{0x1}$. At this stage, the coefficient pairs are structured and loaded into memory, ready for computation.

   \malik{Stage-1 of the FNTT begins by loading paired coefficients from memory and processing them using adjacent pairing operations. Results are written back to the same addresses, enabling efficient reuse. All four DP-BRAMs operate concurrently, maximizing parallelism. In stage-2, the output is reloaded with an offset of 2, coefficients are swapped, and processed outputs are stored at the same addresses, maintaining full utilization. In stage-3, the offset reduces to 1, which is smaller than $\mathrm{\frac{8\text{-point NTT}}{4\text{-memories}} = 2}$ addresses per memory. Initially, $\texttt{DP-BRAM-1}$ and $\texttt{DP-BRAM-2}$ are active, while $\texttt{DP-BRAM-3}$ and $\texttt{DP-BRAM-4}$ remain idle. Once processed, the remaining two memories become active, continuing this alternating pattern until all addresses are covered.}

   \malik{A key observation in stage-3 is that the final FNTT coefficients are scattered across $\texttt{DP-BRAM-1}$ to $\texttt{DP-BRAM-4}$ rather than stored sequentially. As shown in Fig.~\ref{fig:8_point_ntt_cstudy}, the colored boxes (green, orange, blue, red) highlight this distribution. Specifically, coefficients appear at address $\mathrm{0x0}$ in $\texttt{DP-BRAM-1}$ and $\texttt{DP-BRAM-3}$, and at $\mathrm{0x1}$ in $\texttt{DP-BRAM-2}$ and $\texttt{DP-BRAM-4}$. For Kyber, the final coefficients emerge after stage-2 and require reordering, which can be performed using Fig.~\ref{fig:coefficients_reoredring}. The INTT follows the reverse flow of the FNTT.}

    %%==================================%%
    % Section 5: Proposed NTT Accelerator Design
    %%==================================%%
    
    \section{Proposed Accelerator Architecture} \label{sec:proposed_architecture}

\begin{figure}[tb]
    \centering
        % \vspace{-2.5mm}
        \includegraphics[width=\columnwidth]{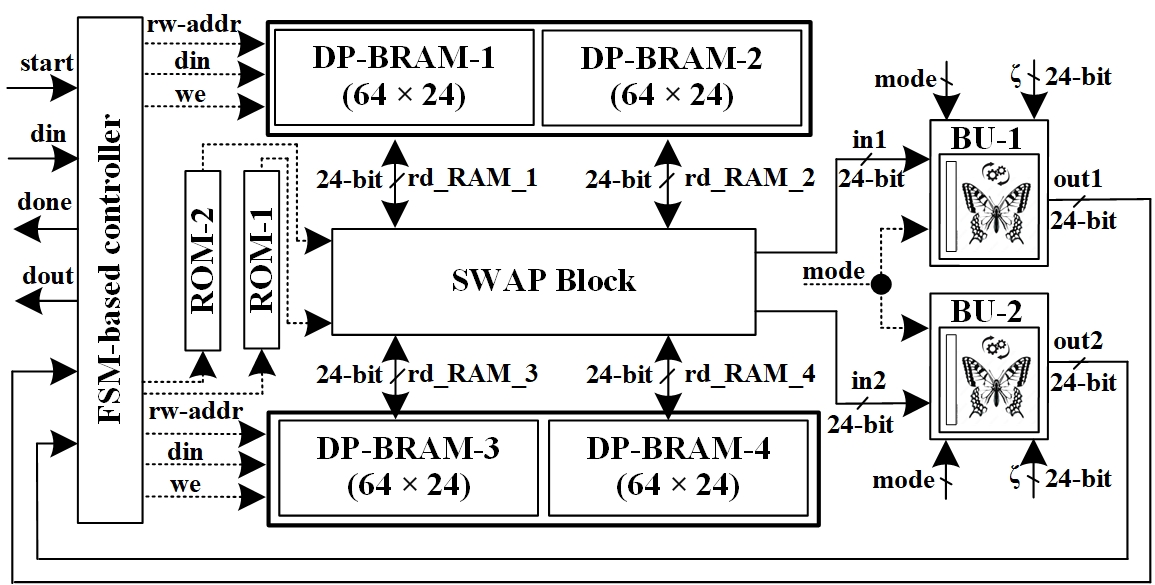}
        \caption{PIP-NTT: The proposed pipelined NTT design.   
        }
        % \vspace{-2.5mm}
    \label{fig:PIP_NTT}
    \end{figure}

\begin{figure}[tb]
    \centering
        \includegraphics[width=\columnwidth]{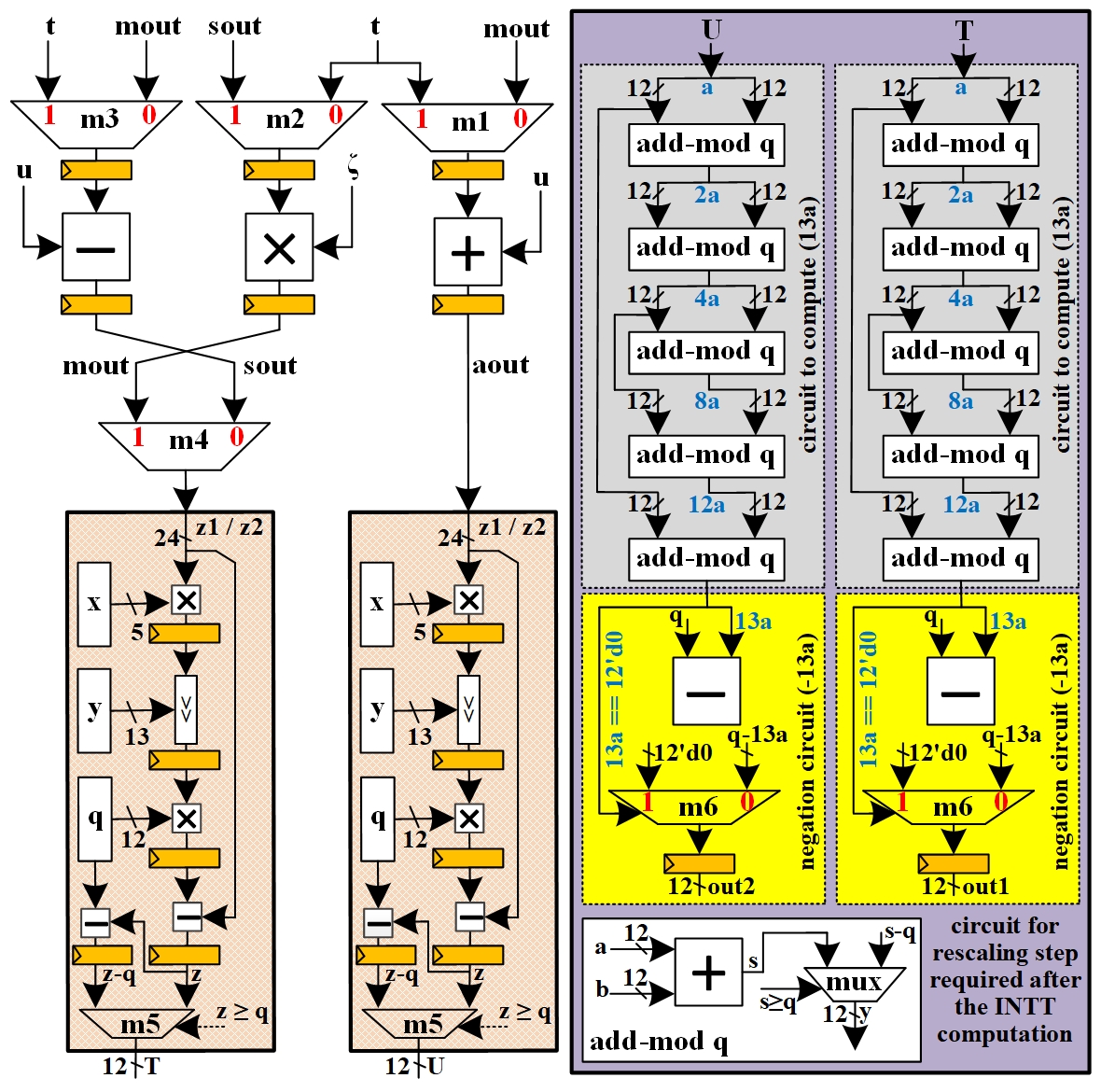}
        %\vspace{-3.0mm}
        \caption{Proposed butterfly unit architecture. \malik{Pattern-filled boxes represent Barrett-based modular reduction units used to compute the $\texttt{mod~q}$ operations in both Cooley-Tukey and Gentleman-Sande configurations. The right portion of the figure implements the rescaling step $a \cdot n^{-1}~\texttt{mod q}$ as described in line 23 of Algorithm~\ref{alg:fntt}. The circuits highlighted in gray and yellow correspond to the computation of $13a$ and its conditional negation $-13a$, respectively.}       }   
        
        % \vspace{-7mm}
    \label{fig:butterfly}
    \end{figure}

    \malik{The PIP-NTT accelerator in Fig.~\ref{fig:PIP_NTT} combines the radix-2 optimized \textsc{8SP-BU} unit with memory-parallelization to implement Algorithm~\ref{alg:fntt} for FNTT and INTT. It includes two butterfly units (BU-1 and BU-2), four memory blocks, two ROMs for twiddle factors, a swap unit for operand reordering, and an FSM controller. The butterfly units, detailed in Section~\ref{sec:dse}, handle arithmetic and modular reduction. To complete the pipeline, the rescaling step (line 23 of Algorithm~\ref{alg:fntt}) is integrated into the butterfly datapath, as shown in Fig.~\ref{fig:butterfly}.}

    %%==================================%%
    % Figure for PIP-NTT
    %%==================================%%

    %%==================================%%
    % Optimized Butterfly Units
    %%==================================%%

    %%==================================%%
    % Figure for Butterfly Unit
    %%==================================%%

    \subsection{Integrated Butterfly and Rescaling Architecture} \label{subsec:rescaling_unit}
    
    \malik{Fig.~\ref{fig:butterfly} shows the integrated butterfly and rescaling unit. The left side replicates the \textsc{8SP-BU} architecture from Fig.~\ref{fig:8SP}, with added pipeline registers after multiplexer \texttt{m4} and the adder. The right side implements line 23 of Algorithm~\ref{alg:fntt}, which performs the rescaling step after inverse NTT. Instead of costly division by $n$, we multiply each coefficient by $n^{-1} \mod \texttt{q}$, where $n$ is the ring size and $\texttt{q}$ is the modulus. This modular inverse-based approach is optimized for Kyber parameters and detailed below.}

    \subsubsection{\muhammad{Rescaling optimization via modular inverse}}

    \malik{
    For~Kyber, with $n = 256$ and $\texttt{q} = 3329$, the modular inverse $n^{-1} = 3316$ is precomputed. A shift-add circuit using its binary form requires seven additions, increasing slice and LUT usage. Instead, we use the congruence $3316 \equiv -13 \mod 3329$, enabling rescaling as $y = (-13 \cdot a) \mod 3329$. Since $13 = 8 + 4 + 1$, the value $13a$ is computed using minimal hardware cost.}

    \begin{align}
    2a  = (a+a)~\texttt{mod q} \nonumber\\
    4a  = (2a+2a)~\texttt{mod q} \nonumber\\
    8a  = (4a+4a)~\texttt{mod q} \nonumber\\
    13a = (8a+4a+a)~\texttt{mod q} \nonumber\\
    y = \begin{cases}
        0 & 13a=0\\
        \texttt{q}-13a & \text{otherwise}
    \end{cases}
    \label{eq:our_const_mul}
    \end{align}

    \subsubsection{\muhammad{Lightweight rescaling via modular doublings}}

    \malik{
    In Eq.~\ref{eq:our_const_mul}, $2a$, $4a$, and $8a$ are computed via modular doublings (each using an addition followed by $\texttt{mod q}$). To form $13a$, we combine $8a$, $4a$, and $a$ using a ternary composition. This can be reorganized as two steps: compute $12a = 8a + 4a$, then $13a = 12a + a$. In total, five modular doublings are needed. To obtain $-13a$, a conditional subtraction is applied, as shown in Eq.~\ref{eq:our_mod_q}, where each modular addition checks if the sum exceeds $\texttt{q}$ and subtracts accordingly.
    }

    \begin{equation}
    s=a+b, s' = s-\texttt{q}, y'= \big(s\ge \texttt{q}\big)\ ?\ s' : s
    \label{eq:our_mod_q}
    \end{equation}

     \subsubsection{\muhammad{Resource and performance comparison}}
     \malik{
     The right portion of Fig.~\ref{fig:butterfly} implements Eq.~\ref{eq:our_const_mul} and Eq.~\ref{eq:our_mod_q}, which together realize the rescaling step using modular addition. Compared to the 7-term shift-add circuit, our constant multiplication circuit consumes 1.67$\times$ fewer slices and 1.84$\times$ fewer LUTs, with a 1.17$\times$ increase in FFs due to local buffering, while producing an identical 12-bit result. 
    The operating frequency remains nearly unchanged, measured at 200 MHz for the shift-add circuit and 204 MHz for our modular addition-based implementation.
    While the rescaling unit finalizes the arithmetic pipeline by efficiently computing $a \cdot n^{-1}~\texttt{mod q}$, its performance is complemented by the underlying memory architecture that sustains high throughput. To support concurrent operation of the two butterfly units and maintain a steady data supply, the PIP-NTT design employs a parallel memory subsystem, described below.}

    %%==================================%%
    % Memory Blocks
    %%==================================%%
    
    \subsection{Dual-Port BRAMs + ROMs}

    As shown in Fig.~\ref{fig:PIP_NTT}, to use BU-1 and BU-2 in parallel, our PIP-NTT design incorporates four dual port BRAMs ($\texttt{DP-BRAM-1}$ to $\texttt{DP-BRAM-4}$), each sized at $\frac{n}{4} \times 24$ bits. Here, $\frac{n}{4}$ denotes the total number of memory addresses per BRAM, while 24 bits correspond to a pair of two 12 bit Kyber coefficients stored in each address. In addition, PIP-NTT uses $\texttt{ROM-1}$ and $\texttt{ROM-2}$ to store precomputed twiddle factors $\zeta$ for FNTT and INTT, respectively. Each ROM unit has a size of $128 \times 24$, where each of the 128 addresses holds two 12 bit twiddle factors. The corresponding $\zeta$ values are forwarded to BU-1 and BU-2 for computation.

    %%==================================%%
    % Swap Unit
    %%==================================%%

    \subsection{Swap Block} \label{subsec:swap_unit_CU}
    \malik{The swap unit in the PIP-NTT architecture is placed between the memory blocks and butterfly units, as shown in Fig.~\ref{fig:PIP_NTT}. It receives coefficient pairs from four DP-BRAMs via $\texttt{rd\_RAM\_1}$ to $\texttt{rd\_RAM\_4}$, reorders them based on the current stage, and routes the correctly paired operands to the butterfly blocks. Implemented purely as a routing network using multiplexers, the swap unit requires no extra BRAM, RegFile, or FIFO storage and operates without additional clock cycles. Although not shown in Fig.~\ref{fig:PIP_NTT}, these multiplexers dynamically form operand pairs for each stage. The swap logic supports both adjacent-coefficient processing and offset-based swapping, as detailed in Sections~\ref{subsubsec:adjacent_coefficient_processing} and~\ref{subsubsec:swapping}.}

    %%==================================%%
    % FSM Controller
    %%==================================%%

    \subsection{FSM Controller and Clock Cycle Calculation}
    \malik{The FSM controller orchestrates Algorithm~\ref{alg:fntt} and the architecture in Fig.~\ref{fig:PIP_NTT} by managing loop counters, generating read/write signals for $\texttt{DP-BRAM-1}$ to $\texttt{DP-BRAM-4}$ based on step size $\delta$, and issuing twiddle factor reads from $\texttt{ROM-1}$ and $\texttt{ROM-2}$. It also controls the routing multiplexers in Fig.~\ref{fig:PIP_NTT} and Fig.~\ref{fig:butterfly} to ensure correct operand alignment. After both forward and inverse NTT, it reorders the coefficients to their natural order using address generation logic, as shown in Fig.~\ref{fig:coefficients_reoredring}.}

    %%==================================%%
    % Cycle Counts
    %%==================================%%

    % \subsection{Clock Cycle Calculation}
    \malik{With four $\frac{n}{4}$-sized memory units and two butterfly units operating in parallel, PIP-NTT completes one forward and one inverse NTT operation in 512 clock cycles, calculated as $\mathrm{log}_2(n) \times 64$. In each stage, four 12-bit coefficients are processed over 64 cycles. The \textsc{8SP-BU} units contribute an additional 8 cycles due to their internal 8-stage pipelining.}

    %%=====================================
    %% Section 6: Results and Comparisons
    %%=====================================
  
    \section{Implementation Results and Comparisons}\label{sec:results_and_comparisons}

    All the architectures of Fig.~\ref{fig:overall}, Fig.~\ref{fig:PIP_NTT} and Fig.~\ref{fig:butterfly} were implemented in Verilog HDL using the \texttt{Vivado IDE (v2023.2)}. The hardware area results for the design space of the butterfly units were evaluated in terms of FPGA slices, LUTs, and FFs, excluding DSP blocks. In contrast, DSPs are also considered to evaluate the hardware area of the PIP-NTT. 

    %%=====================================
    %% Implementation Results
    %%=====================================
    
    \subsection{Implementation Results}\label{subsec:res_1}

    %Fig.~\ref{fig:dpath_delay} summarizes the data path delay of the non-pipelined and pipelined butterfly designs. It shows that pipelining reduces the data path delay when moving from \textsc{BL-BU} to the most optimized \textsc{8SP-BU}. The red-dashed lines in Fig.~\ref{fig:dpath_delay} illustrate the benefits of pipelining in modular reduction operations, as the critical path delay has been significantly reduced. This signifies the importance of pipelined reduction methods when implementing the NTT for PQC algorithms.

\begin{figure}[tb]
    \centering
    \begin{tikzpicture}
        \begin{axis}[
        legend style={draw=none},   
        ybar,
        every node near coord/.append style={rotate=90, anchor=west},
        bar width = 5.5pt,
        enlarge x limits=0.24, % Adjust for spacing
        ymax=40,
        ylabel={Delay (ns)}, % Added unit in ns
        ylabel style={yshift=-7pt}, % Adjusted to reduce space between ylabel and y-axis
        symbolic x coords={TD, RD, LD},
        xtick=data,
        nodes near coords,
        width = 8.25cm,
        height = 5.0cm,
        legend style={at={(1.0,0.45)}, anchor=west,legend columns=1},
        legend image post style = {scale=0.7},
        font=\footnotesize,
        ]
            \addplot[ybar=black, fill=red] coordinates {(TD, 23.953) (RD, 13.740) (LD, 10.213)};
            \addplot[ybar=black, fill=blue] coordinates {(TD, 10.750) (RD, 6.363) (LD, 4.387)};
            \addplot[ybar=black, fill=green] coordinates {(TD, 6.817) (RD, 3.397) (LD, 3.420)};
            \addplot[ybar=black, fill=violet] coordinates {(TD, 6.206) (RD, 2.855) (LD, 3.351)};
            \addplot[ybar=black, fill=blue!25] coordinates {(TD, 3.979) (RD, 2.177) (LD, 1.802)};
            \addplot[ybar=black, fill=teal] coordinates {(TD, 3.571) (RD, 1.733) (LD, 1.838)};
            \addplot[ybar=black, fill=purple] coordinates {(TD, 3.511) (RD, 1.728) (LD, 1.783)};
            \addplot[ybar=black, fill=gray] coordinates {(TD, 3.270) (RD, 1.623) (LD, 1.647)};
            \legend {\textsc{BL-BU}, \textsc{2SP-BU}, \textsc{3SP-BU}, \textsc{4SP-BU}, \textsc{5SP-BU}, \textsc{6SP-BU}, \textsc{7SP-BU}, \textsc{8SP-BU}};

            \draw[dashed,red] (1,-30) rectangle (50,250);
            \draw[dotted,black] (33,-50) rectangle (47,230);

            \draw[red,->] (40,320) -- (20,275);
            \draw[dotted,black,->] (70,285) -- (45,200);

            \node[red] at (100,350) {Pipelining (mod~q) operation optimizes NTT frequency};
            \node[black] at (95,295) {Optimized butterfly};
            
        \end{axis}
    \end{tikzpicture}
    % \vspace{-2mm}
    \caption{Data path delay values. TD, RD, and LD represent the total data path, routing, and logic delays, respectively.}
    % \vspace{-2.5mm}
    \label{fig:dpath_delay}
\end{figure}
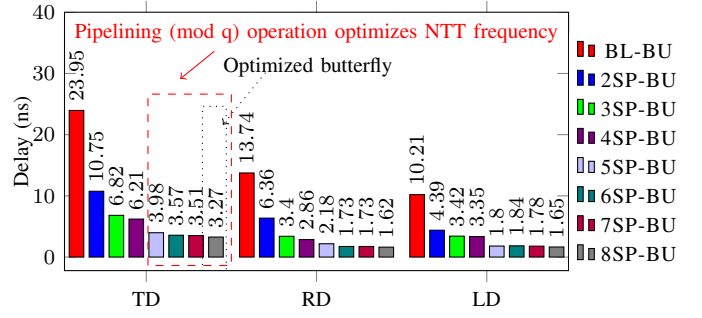

    Fig.~\ref{fig:dpath_delay} compares data path delays for non-pipelined and pipelined butterfly designs. It highlights how pipelining (from \textsc{BL-BU} to the optimized \textsc{8SP-BU}) significantly reduces critical path delay, especially in modular reduction. The red-dashed lines emphasize the impact of pipelined reduction, underscoring its importance for efficient NTT implementation in PQC systems.

    %%=====================================
    %% Data Path Delay Figure
    %%=====================================
    
  %  Table~\ref{tab:results} provides the hardware area and timing results for the butterfly architectures (after post-place and route) of Fig.~\ref{fig:overall}, where the computation time (i.e., Time in Table~\ref{tab:results}) for one FNTT and INTT operations is calculated using Eq.~\ref{eq:computation_time}. It reveals that increasing the number of pipeline stages leads to an increase in both the hardware area and the operating frequency, while the computation time decreases. To illustrate this, let us compare butterfly architectures without pipelined modular reduction operations (from \textsc{BL-BU} to \textsc{4SP-BU}). The \textsc{4SP-BU} design, which is the most optimized in terms of operating frequency, shows a significant 3.93$\times$ increase in circuit frequency compared to \textsc{BL-BU}, with only a 1.15$\times$ increase in slice area. When comparing \textsc{4SP-BU} to the most optimized \textsc{8SP-BU}, there is a 1.64$\times$ increase in circuit frequency with a 1.14$\times$ increase in slice area. Although the increases in circuit frequency and area overhead are similar, the operating frequency still improves. The most notable trade-off is observed when comparing \textsc{BL-BU} with the most optimized \textsc{8SP-BU}, which shows a 6.47$\times$ increase in circuit frequency with only a 1.32$\times$ increase in slice area. This highlights the significant performance gains achieved through pipelining at a fine-grained level, despite the modest increase in hardware area.

  Table~\ref{tab:results} summarizes post-place-and-route area and timing for the butterfly designs in Fig.~\ref{fig:overall}, with computation time derived from Eq.~\ref{eq:computation_time}. Increasing pipeline stages boosts both area and frequency, while reducing computation time. For example, \textsc{4SP-BU} achieves a 3.93$\times$ frequency gain over \textsc{BL-BU} with only 1.15$\times$ more slice area. Compared to \textsc{4SP-BU}, the fully optimized \textsc{8SP-BU} offers a further 1.64$\times$ frequency increase with just 1.14$\times$ more area. The largest gain is from \textsc{BL-BU} to \textsc{8SP-BU}, showing a 6.47$\times$ frequency boost with only 1.32$\times$ area overhead. It highlights the efficiency of fine-grained pipelining.

    \begin{equation}\label{eq:computation_time}
        \text{Time}~(\mu s) = \frac{\text{Clock Cycles (CCs)}}{\text{Frequency (in MHz)}}
    \end{equation}

    %%=====================================
    %% Table for butterfly unit results
    %%=====================================

    \begin{table}[tb]
    %\centering
    \caption{Results for butterfly units on Virtex-7 FPGA with $n=256$ and $\texttt{q}=3329$. The \textbf{\cmark} shows modular reduction is pipelined (and indicates better performance in frequency).}
    %\vspace{-2.5mm}
    %\begin{center}
    \begin{tabular}{|c|c|c|c|c|c|c|}
    \hline
    
    \multirow{3}{*}{\textbf{Designs}} &
    \multicolumn{3}{l|}{\textbf{Hardware Area}} &
    \multicolumn{3}{l|}{\textbf{Timing Results}} \\ \cline{2-7}

    {} & 
    \multirow{2}{*}{\textbf{Slices}} & 
    \multirow{2}{*}{\textbf{LUTs}} & 
    \multirow{2}{*}{\textbf{FFs}} & 
    \multirow{1}{*}{\textbf{Latency}} & 
    \textbf{Frequency} & 
    \textbf{Time} \\ 

    {} & 
    {} & 
    {} & 
    {} & 
    {\textbf{(CCs)}} & 
    \textbf{(MHz)} & 
    \textbf{($\mu s$)} \\ \hline \hline

    % \multicolumn{7}{|l|}{\textbf{\textcolor{black}{Results for the butterfly units of Fig.~\ref{fig:overall}}}} \\ \hline \hline

    {\textsc{BL-BU}} & {231} & {713} & {65} & {900} & {46} & {19.56} \\ %\hline
    {\textsc{2SP-BU}} & {243} & {722} & {112} & {903} & {80} & {11.28}  \\ %\hline
    {\textsc{3SP-BU}} & {260} & {761} & {182} & {904} & {133} & {6.79} \\ %\hline
    {\textsc{4SP-BU}} & {267} & {718} & {229} & {905} & {181} & {5.00} \\ %\hline
    {\textsc{5SP-BU}} & {263} & {714} & {253} & {906} & {222 \cmark} & {4.08} \\ %\hline
    {\textsc{6SP-BU}} & {260} & {725} & {277} & {907} & {285 \cmark} & {3.18} \\ %\hline
    {\textsc{7SP-BU}} & {285} & {724} & {301} & {908} & {296 \cmark} & {3.06} \\ %\hline
    {\textsc{8SP-BU}} & {306} & {873} & {347} & {909} & {298 \cmark} & {3.05} \\ \hline

    \end{tabular}
    \footnotesize{\\We considered up to 8-pipeline stages, as more stages only increase the area without further optimizing frequency. This is evident where \textsc{7SP-BU} and \textsc{8SP-BU} operate on a maximum of 296 MHz and 298 MHz, respectively, while these architectures have a significant area difference.}
    \label{tab:results}
    \end{table}

    The area and timing results for two versions of PIP-NTT are summarized in Table~\ref{tab:results_PIP_NTT}. Version-I is a pipelined accelerator, which uses three RegBanks and an \textsc{8SP-BU} butterfly unit of~Fig.~\ref{fig:8SP}. Two RegBanks store coefficients for ping-pong processing, while the third holds twiddle factors. Version-II is a pipelined parallel NTT accelerator of~Fig.~\ref{fig:PIP_NTT}.

    %%=====================================
    %% Table for PIP-NTT results
    %%=====================================

    \begin{table}[tb]
    \centering
    \caption{Results for PIP-NTT on Virtex-7 FPGA.}
    %\vspace{-2.5mm}
    %\begin{center}
    \begin{tabular}{|c|c|c|c|c|c|c|}
    \hline
    
    \multirow{3}{*}{\textbf{\rotatebox{90}{Work}}} &
    \multicolumn{3}{l|}{\textbf{Hardware Area}} &
    \multicolumn{3}{l|}{\textbf{Timing Results}} \\ \cline{2-7}

    {} & 
    \multirow{2}{*}{\textbf{Slices}} & 
    \multirow{2}{*}{\textbf{LUTs}} & 
    \multirow{2}{*}{\textbf{FFs}} & 
    \multirow{1}{*}{\textbf{Latency}} & 
    \textbf{Frequency} & 
    \textbf{Time} \\ 

    {} & 
    {} & 
    {} & 
    {} & 
    {\textbf{(CCs)}} & 
    \textbf{(MHz)} & 
    \textbf{($\mu s$)} \\ \hline \hline

    \multirow{4}{*}{\rotatebox{90}{PIP-NTT}} & \multicolumn{6}{l|}{\textbf{\textcolor{black}{Version-I (three RegBanks + one \textsc{8SP-BU} of Fig.~\ref{fig:8SP})}}} \\ \cline{2-7} 
    
    {} & \multirow{1}{*}{1681} & \multirow{1}{*}{5939} & \multirow{1}{*}{3877} & \multirow{1}{*}{909} & \multirow{1}{*}{295} & \multirow{1}{*}{3.08} \\ \cline{2-7}

    {} & \multicolumn{6}{l|}{\textbf{\textcolor{black}{Version-II (NTT accelerator of Fig.~\ref{fig:PIP_NTT})}}} \\ \cline{2-7}
    
    {} & \multirow{1}{*}{\textcolor{black}{156}} & \multirow{1}{*}{\textcolor{black}{426}} & \multirow{1}{*}{\textcolor{black}{379}} & \multirow{1}{*}{520} & \multirow{1}{*}{200} & \multirow{1}{*}{2.60} \\ \hline %\cline{5-5} \cline{7-7}

    % {} & \multirow{1}{*}{311} & \multirow{1}{*}{979} & \multirow{1}{*}{323} & \multirow{1}{*}{520} & \multirow{1}{*}{200} & \multirow{1}{*}{2.60} \\ \hline %\cline{5-5} \cline{7-7}

    %{} & {} & {} & {} & {1549} & {} & {5.25} & {--} \\ \hline

    \end{tabular}
    \vspace{-1.5mm}
    \label{tab:results_PIP_NTT}
    %\vspace{-5mm}
    \end{table}

    Table~\ref{tab:results_PIP_NTT} reveals that version-II of PIP-NTT is more efficient in terms of area and computation time than version-I. In version-I, RegBanks (including the FSM controller) use 82\% of slices, while \textsc{8SP-BU} of Fig.~\ref{fig:8SP} accounts for 18\%. Moreover, version-I achieves a 295 MHz frequency, only 3 MHz (1.01\%) lower than the standalone \textsc{8SP-BU} (298 MHz shown in Table~\ref{tab:results}), derived via DSE. \malik{Version-II of PIP-NTT operates at a 1.47$\times$ lower circuit frequency but achieves 10.77$\times$ and 13.94$\times$ higher slice and LUT efficiency, respectively, and 1.18$\times$ faster computation. However, from then on, all subsequent discussions focus on the most efficient PIP-NTT version-II accelerator, which is simply denoted PIP-NTT. 
    }
    %%==================================%%
    % Comparisons to State-of-the-Art
    %%==================================%%

     \subsection{Comparison With Related Works} \label{subsec:comparisons}
    
    Our PIP-NTT is compared with existing NTT designs in Table~\ref{tab:comparisons} using identical devices as used in state-of-the-art implementations. For a fair evaluation, the slice equivalent cost (SEC) is used as the area metric, computed as \#BRAMs $\times$ 200 + \#DSPs $\times$ 100 + \#Slices, following the approximation method of~\cite{SEC_calculation, ref_for_SEC}, where one DSP and one 36Kb BRAM correspond to 102.4 and 196.2 slices, respectively. The ATP is defined as SEC $\times$ Time, with Time representing the computation cost (in $\mu s$) for one FNTT and INTT operation.

    %%=====================================
    %% Table for comparisons
    %%=====================================
    \begin{table*}[!]
    \caption{
    Comparison to FNTT and INTT prior accelerators. All these use $n=256$ and $\texttt{q}=3329$. The \textcolor{blue}{$\uparrow$} and \textcolor{red}{$\downarrow$} correspond to superior and inferior performance of version-II of PIP-NTT compared to reference works, respectively. The ``--'' symbol shows where the relevant information is not available in the reference works.}

    %\vspace{-2.5mm}
    %\begin{center}
    \begin{tabular}
    {|c|c|c|c|c|c|c|c|c|c|c|c|c|} 
    \hline

    \multirow{2}{*}{\textbf{Ref.~\# / Year}} &
    \multirow{2}{*}{\textbf{Device}} &
    \multirow{1}{*}{\textbf{NTT}} &
    \multirow{2}{*}{\textbf{Slices$^1$}} & 
    \multirow{2}{*}{\textbf{LUTs}} & 
    \multirow{2}{*}{\textbf{FFs}} & 
    \multirow{2}{*}{\textbf{DSPs}} & 
    \multirow{2}{*}{\textbf{BRAMs}} & 
    \textbf{Freq} & 
    \multirow{1}{*}{\textbf{Lat}} & 
    \multirow{1}{*}{\textbf{Time}} &
    \multirow{2}{*}{\textbf{SEC$^2$}} &
    \multirow{1}{*}{\textbf{ATP}} \\ 

    {} & 
    {} & 
    {\textbf{Op}} & 
    {} & 
    {} & 
    {} & 
    {} & 
    {} &
    {\textbf{(MHz)}} & 
    {\textbf{(CCs)}} &
    \textbf{($\mu s$)} &
    {} &
    {(SEC $\times$ $\mu s$)} \\ \hline \hline

    \multicolumn{13}{|l|}{\textbf{FNTT and INTT accelerators (allowing computation of the PWM using the same butterfly unit) }} \\ \hline \hline

    %%%%%%%%%%%%%%%%%%%%%%%%%%%%%%%%%%%%
    % reference 7 (Artix-7)
    %%%%%%%%%%%%%%%%%%%%%%%%%%%%%%%%%%%

    \multirow{2}{*}{\cite{ntt_comparison} / 2021} &
    \multirow{2}{*}{{Artix-7}} &
    {FNTT} & 
    \multirow{2}{*}{{187}} &
    \multirow{2}{*}{{360}} & 
    \multirow{2}{*}{{145}} &
    \multirow{2}{*}{{3}} &
    \multirow{2}{*}{{2 (36KB)}} & 
    \multirow{2}{*}{{115}} & 
    {940} & 
    {8.17} & \multirow{2}{*}{887} & {7247 (2.34$\times$\textcolor{blue}{$\uparrow$})} \\ \cline{3-3} \cline{10-11} \cline{13-13}
    {} & {} & {INTT} & {} & {} & {} & {} & {} & {} & {1203} & {10.46} & {} & {9278 (3.00$\times$\textcolor{blue}{$\uparrow$})} \\ \cline{1-13} 

    %%%%%%%%%%%%%%%%%%%%%%%%%%%%%%%%%%%%
    % reference 10 (Virtex-7)
    %%%%%%%%%%%%%%%%%%%%%%%%%%%%%%%%%%%

    \multirow{2}{*}{\cite{COHA_NTT} / 2022} & 
    \multirow{2}{*}{Virtex-7} &
    {FNTT} & 
    \multirow{2}{*}{{675 $^*$}} &
    \multirow{2}{*}{{2128}} & 
    \multirow{2}{*}{{1144}} & 
    \multirow{2}{*}{{8}} &
    \multirow{2}{*}{{3 (36KB)}} &
    \multirow{2}{*}{174} & 
    \multirow{1}{*}{{922}} & 
    {5.29} & \multirow{2}{*}{2075} & {10977 (3.55$\times$\textcolor{blue}{$\uparrow$})}\\ \cline{3-3} \cline{10-11} \cline{13-13}
    {} & {} & {INTT} & {} & {} & {} & {} & {} & {} & {1184} & {6.80} & {} & {14110 (4.57$\times$\textcolor{blue}{$\uparrow$})} \\ \cline{1-13}

    %%%%%%%%%%%%%%%%%%%%%%%%%%%%%%%%%%%%
    % reference 8 (Artix-7)
    %%%%%%%%%%%%%%%%%%%%%%%%%%%%%%%%%%%

    \multirow{2}{*}{\cite{ntt_kyber_16BTF} / 2021} & 
    \multirow{2}{*}{{Artix-7}} &
    {FNTT} & 
    \multirow{2}{*}{{2713 $^*$}} &
    \multirow{2}{*}{{9508}} & 
    \multirow{2}{*}{{2684}} & 
    \multirow{2}{*}{{16}} &
    \multirow{2}{*}{{35 (36KB)}} &
    \multirow{2}{*}{{172}} & 
    {69} & 
    {0.40} &
    \multirow{2}{*}{11313} & 
    {4525 (1.46$\times$\textcolor{blue}{$\uparrow$})} \\ \cline{3-3} \cline{10-11} \cline{13-13}

    {} & {} & {INTT} & {} & {} & {} & {} & {} & {} & {71} & {0.41} & {} & {4638 (1.50$\times$\textcolor{blue}{$\uparrow$})} \\ \hline 

    %%%%%%%%%%%%%%%%%%%%%%%%%%%%%%%%%%%%
    % AREA_TIME_ML_KEM_4BTU_2025
    %%%%%%%%%%%%%%%%%%%%%%%%%%%%%%%%%%%

    \multirow{2}{*}{\cite{AREA_TIME_ML_KEM_4BTU_2025} / \malik{2025}} & 
    \multirow{2}{*}{{\malik{Artix-7}}} &
    {FNTT} & 
    \multirow{2}{*}{{\malik{815}}} &
    \multirow{2}{*}{{\malik{2313}}} & 
    \multirow{2}{*}{{\malik{1630}}} & 
    \multirow{2}{*}{{\malik{4}}} &
    \multirow{2}{*}{{\malik{0}}} &
    \multirow{2}{*}{{\malik{--}}} & 
    \multirow{2}{*}{\malik{--}} & 
     \multirow{2}{*}{\malik{--}} &
    \multirow{2}{*}{\malik{1215}} & 
     \multirow{2}{*}{\malik{--}} \\ \cline{3-3}
    
    {} & {} & {INTT} & {} & {} & {} & {} & {} & {} & {} & {} & {} & {} \\ \hline \hline

    \multicolumn{13}{|l|}{\textbf{Accelerators supporting only the FNTT and INTT operations (without the PWM) \cite{HyperNTT, poly_mult_acc_for_K_2024, K_NTT_64_BTFs_2024, pipeline_ntt, CRYPHTOR, K_NTT_only_AREA_2024, sajjad_access_ntt} }} \\ \hline \hline

    \multicolumn{13}{|l|}{\textbf{NTT accelerators prioritizing high-speed in computation time and efficient in low-cycle counts}} \\ \hline \hline

    %%%%%%%%%%%%%%%%%%%%%%%%%%%%%%%%%%%%
    % HyperNTT (Artix-7)
    %%%%%%%%%%%%%%%%%%%%%%%%%%%%%%%%%%%

    \multirow{2}{*}{\cite{HyperNTT} / 2024} & 
    \multirow{2}{*}{Artix-7} &
    {FNTT} & 
    \multirow{2}{*}{{1626}} &
    \multirow{2}{*}{{5156}} & 
    \multirow{2}{*}{{6606}} & 
    \multirow{2}{*}{{7}} &
    \multirow{2}{*}{{4 (36KB)}} &
    \multirow{2}{*}{393} & 
    \multirow{1}{*}{{84}} & 
    {0.21} & \multirow{2}{*}{3126} & {656 (4.70$\times$\textcolor{red}{$\downarrow$})}\\ \cline{3-3} \cline{10-11} \cline{13-13}
    
    {} & {} & {INTT} & {} & {} & {} & {} & {} & {} & {84} & {0.21} & {} & {656 (4.70$\times$\textcolor{red}{$\downarrow$})} \\ \hline

    %%%%%%%%%%%%%%%%%%%%%%%%%%%%%%%%%%%%
    % reference 19 (Artix-7)
    %%%%%%%%%%%%%%%%%%%%%%%%%%%%%%%%%%%

    \multirow{2}{*}{\cite{poly_mult_acc_for_K_2024} / 2023} & 
    \multirow{2}{*}{{Artix-7}} &
    {FNTT} & 
    \multirow{2}{*}{{416}} &
    \multirow{2}{*}{{1170}} & 
    \multirow{2}{*}{{1164}} & 
    \multirow{2}{*}{{4}} &
    \multirow{2}{*}{{2 (36KB)}} &
    \multirow{2}{*}{{303}} & 
    {235} & 
    {0.78} &
    \multirow{2}{*}{1216} & 
    {948 (3.25$\times$\textcolor{red}{$\downarrow$})} \\ \cline{3-3} \cline{10-11} \cline{13-13}
    
    {} & {} & {INTT} & {} & {} & {} & {} & {} & {} & {235} & {0.78} & {} & {948 (3.25$\times$\textcolor{red}{$\downarrow$})} \\ \hline

    % %%%%%%%%%%%%%%%%%%%%%%%%%%%%%%%%%%%%
    % % K_NTT_64_BTFs_2024
    % %%%%%%%%%%%%%%%%%%%%%%%%%%%%%%%%%%%
    
    \multirow{2}{*}{\cite{K_NTT_64_BTFs_2024} / 2024} & 
    \multirow{2}{*}{{Artix-7}} &
    {FNTT} & 
    \multirow{2}{*}{{6091}} &
    \multirow{2}{*}{{18296}} & 
    \multirow{2}{*}{{12134}} & 
    \multirow{2}{*}{{64}} &
    \multirow{2}{*}{{0}} &
    \multirow{2}{*}{{210}} & 
    {85} & 
    {0.40} &
    \multirow{2}{*}{12491} & 
    {4996 (1.61$\times$\textcolor{blue}{$\uparrow$})} \\ \cline{3-3} \cline{10-11} \cline{13-13}
    
    {} & {} & {INTT} & {} & {} & {} & {} & {} & {} & {104} & {0.50} & {} & {6245 (2.02$\times$\textcolor{blue}{$\uparrow$})} \\ \hline

    % %%%%%%%%%%%%%%%%%%%%%%%%%%%%%%%%%%%%
    % % high_speed_NTT_kyber_2024
    % %%%%%%%%%%%%%%%%%%%%%%%%%%%%%%%%%%%
    
    \multirow{1}{*}{\cite{high_speed_NTT_kyber_2024} / \malik{2024}} & 
    \multirow{1}{*}{{\malik{Artix-7}}} &
    {FNTT} & 
    \multirow{1}{*}{{\malik{401}}} &
    \multirow{1}{*}{{\malik{1070}}} & 
    \multirow{1}{*}{{\malik{1071}}} & 
    \multirow{1}{*}{{\malik{5}}} &
    \multirow{1}{*}{{\malik{10.5 (36KB)}}} &
    \multirow{1}{*}{{\malik{278}}} & 
    {\malik{306}} & 
    {\malik{1.10}} &
    \multirow{1}{*}{\malik{3001}} & 
    {\malik{3301} (\malik{1.07}$\times$\textcolor{blue}{$\uparrow$})} \\ \hline

    \multicolumn{13}{|l|}{\textbf{Other NTT accelerators: prioritizing area efficiency in standalone slices for \cite{CRYPHTOR, K_NTT_only_AREA_2024, iter_ntt_2025_wiley}, in DSPs + BRAMs for \cite{sajjad_access_ntt}}} \\ \hline \hline

    % %%%%%%%%%%%%%%%%%%%%%%%%%%%%%%%%%%%%
    % % reference 9 (Artix-7)
    % %%%%%%%%%%%%%%%%%%%%%%%%%%%%%%%%%%%
    
    \cite{pipeline_ntt} / 2020 & 
    \multirow{1}{*}{{Artix-7}} &
    {FNTT} & 
    {--} & 
    {--} &
    {--} &
    {--} & 
    {--} & 
    {155} & 
    {1834} & 
    {11.83} & {--} & {--} \\ \cline{1-13}

    % %%%%%%%%%%%%%%%%%%%%%%%%%%%%%%%%%%%%
    % % CRYPHTOR
    % %%%%%%%%%%%%%%%%%%%%%%%%%%%%%%%%%%%
    
    \cite{CRYPHTOR} / 2024 & 
    \multirow{1}{*}{{Artix-7}} &
    {FNTT} & 
    {381} & 
    {1243} &
    {562} &
    {11} & 
    {3.5 (36KB)} & 
    {118} & 
    {933} & 
    {7.90} & {2181} & {17229 (5.58$\times$\textcolor{blue}{$\uparrow$})} \\ \cline{1-13}

    % %%%%%%%%%%%%%%%%%%%%%%%%%%%%%%%%%%%%
    % % K_NTT_only_AREA_2024
    % %%%%%%%%%%%%%%%%%%%%%%%%%%%%%%%%%%%
    
    \cite{K_NTT_only_AREA_2024} / 2024 & 
    \multirow{1}{*}{{Artix-7}} &
    {FNTT} & 
    {375} & 
    {1405} &
    {190} &
    {10} & 
    {11 (36KB)} & 
    {262} & 
    {--} & 
    {--} & {3575} & {--} \\ \cline{1-13}

    % %%%%%%%%%%%%%%%%%%%%%%%%%%%%%%%%%%%%
    % % ntt_kyber_1
    % %%%%%%%%%%%%%%%%%%%%%%%%%%%%%%%%%%%

    % \multirow{2}{*}{\cite{ntt_kyber_1} / 2021} & 
    % \multirow{2}{*}{{Artix-7}} &
    % {FNTT} & 
    % \multirow{2}{*}{{232$^*$}} &
    % \multirow{2}{*}{{609}} & 
    % \multirow{2}{*}{{640}} & 
    % \multirow{2}{*}{{2}} &
    % \multirow{2}{*}{{2}} &
    % \multirow{2}{*}{{257}} & 
    % {490} & 
    % {1.90} &
    % \multirow{2}{*}{832} & 
    % {1580 (2.75$\times$\textcolor{red}{$\downarrow$})} \\ \cline{3-3} \cline{10-11} \cline{13-13}
    
    % {} & {} & {INTT} & {} & {} & {} & {} & {} & {} & {490} & {1.90} & {} & {1580 (2.75$\times$\textcolor{red}{$\downarrow$})} \\ \hline 

    %%%%%%%%%%%%%%%%%%%%%%%%%%%%%%%%%%%%
    % sajjad-ntt (Virtex-7)
    %%%%%%%%%%%%%%%%%%%%%%%%%%%%%%%%%%%

    \multirow{2}{*}{\cite{sajjad_access_ntt} / 2024} & 
    \multirow{2}{*}{Virtex-7} &
    {FNTT} & 
    \multirow{2}{*}{{1287}} &
    \multirow{2}{*}{{5109}} & 
    \multirow{2}{*}{{3184}} & 
    \multirow{2}{*}{{0}} &
    \multirow{2}{*}{{0}} &
    \multirow{2}{*}{212} & 
    \multirow{1}{*}{{896}} & 
    {4.22} & \multirow{2}{*}{1287} & {5431 (1.80$\times$\textcolor{blue}{$\uparrow$})}\\ \cline{3-3} \cline{10-11} \cline{13-13}
    
    {} & {} & {INTT} & {} & {} & {} & {} & {} & {} & {1024} & {4.83} & {} & {6216 (2.06$\times$\textcolor{blue}{$\uparrow$})} \\ \hline

    % %%%%%%%%%%%%%%%%%%%%%%%%%%%%%%%%%%%%
    % % reference 9 (Artix-7)
    % %%%%%%%%%%%%%%%%%%%%%%%%%%%%%%%%%%%
    
    \cite{iter_ntt_2025_wiley} / \malik{2025} & 
    \multirow{1}{*}{{\malik{Artix-7}}} &
    {\malik{FNTT}} & 
    {\malik{426}} & 
    {\malik{1264}} &
    {\malik{1000}} &
    {\malik{2}} & 
    {\malik{2 (36KB)}} & 
    {\malik{243}} & 
    {\malik{1034}} & 
    {\malik{4.25}} & 
    {\malik{1026}} & 
    {\malik{4360 (1.41$\times$\textcolor{blue}{$\uparrow$})}} \\ \hline \hline

    %%%%%%%%%%%%%%%%%%%%%%%%%%%%%%%%%%%%
    % safi-ntt
    %%%%%%%%%%%%%%%%%%%%%%%%%%%%%%%%%%%

    % {\cite{safi_ntt}} &
    % {Virtex-7} &
    % {FNTT} & 
    % {1950$^{*}$} &
    % {7800$^\dagger$} &
    % {--} & 
    % {6} &
    % {0} &
    % {72} & 
    % {--} & 
    % {4.04$^\ddagger$} & 
    % {2550} &
    % {10302 (1.98$\times$\textcolor{blue}{$\uparrow$)}} \\ \hline

    %%%%%%%%%%%%%%%%%%%%%%%%%%%%%%%%%%%%
    % reference 14 (Virtex-7)
    %%%%%%%%%%%%%%%%%%%%%%%%%%%%%%%%%%%

    % \multirow{2}{*}{\cite{ESL_malik}} & 
    % \multirow{2}{*}{Virtex-7} &
    % {FNTT} & 
    % \multirow{2}{*}{{3698}} &
    % \multirow{2}{*}{{9298}} & 
    % \multirow{2}{*}{{9402}} & 
    % \multirow{2}{*}{{0}} &
    % \multirow{2}{*}{{0}} &
    % \multirow{2}{*}{20} & 
    % \multirow{1}{*}{{898}} & 
    % {44.90} & \multirow{2}{*}{3698} & {166040 (25.93$\times$\textcolor{blue}{$\uparrow$})}\\ \cline{3-3} \cline{10-11} \cline{13-13}
    % {} & {} & {INTT} & {} & {} & {} & {} & {} & {} & {898} & {44.90} & {} & {166040 (32.07$\times$\textcolor{blue}{$\uparrow$})} \\ \hline \hline

    \multicolumn{13}{|l|}{\textbf{PIP-NTT also supports only the FNTT and INTT operations (without the PWM) }} \\ \hline \hline    
    
    % \multirow{2}{*}{\textbf{TW}} &
    % {Artix-7} &
    % {Both} &
    % {1629} &
    % {5899} &
    % {3879} &
    % {0} &
    % {0} &
    % {231} & 
    % {909} & 
    % {3.93} & {1629} &
    % {6402}\\ \cline{2-13}

    % %%%%%%%%%%%%%%%%%%%%%%%%%%%%%%%%%%%%
    % % PIP-NTT (Virtex-7)
    % %%%%%%%%%%%%%%%%%%%%%%%%%%%%%%%%%%%

    % {} &
    % {Virtex-7} &
    % {Both} & 
    % {1681} &
    % {5939} &
    % {3877} & 
    % {0} &
    % {0} &
    % {295} & 
    % {909} & 
    % {3.08} & 
    % {1681} &
    % {5177} \\ \hline  

    \multirow{4}{*}{\textbf{PIP-NTT}} &
    \multirow{2}{*}{{Artix-7}} &
    \multirow{1}{*}{{FNTT}} &
    \multirow{2}{*}{{\malik{186}}} &
    \multirow{2}{*}{{\malik{531}}} &
    \multirow{2}{*}{{\malik{380}}} &
    \multirow{2}{*}{{6}} &
    \multirow{2}{*}{{4 (18KB)}} &
    \multirow{2}{*}{{200}} & 
    \multirow{2}{*}{{520}} & 
    \multirow{2}{*}{{2.60}} & 
    \multirow{2}{*}{{\malik{1186}}} &
    \multirow{2}{*}{{\malik{3084}}}\\ \cline{3-3}

    {} & {} & {INTT} & {} & {} & {} & {} & {} & {} & {} & {} & {} & {} \\ \cline{2-13}

    {} &
    \multirow{2}{*}{{Virtex-7}} &
    \multirow{1}{*}{{FNTT}} &
    \multirow{2}{*}{{\malik{156}}} &
    \multirow{2}{*}{{\malik{426}}} &
    \multirow{2}{*}{{\malik{379}}} &
    \multirow{2}{*}{{6}} &
    \multirow{2}{*}{{4 (18KB)}} &
    \multirow{2}{*}{{200}} & 
    \multirow{2}{*}{{520}} & 
    \multirow{2}{*}{{2.60}} & 
    \multirow{2}{*}{{\malik{1156}}} &
    \multirow{2}{*}{{\malik{3006}}}\\ \cline{3-3}

    {} & {} & {INTT} & {} & {} & {} & {} & {} & {} & {} & {} & {} & {} \\ \hline

    %%%%%%%%%%%%%%%%%%%%%%%%%%%%%%%%%%%%
    % reference 20 (Artix-7)
    %%%%%%%%%%%%%%%%%%%%%%%%%%%%%%%%%%%

    % \multirow{2}{*}{\cite{Ziying_2023} / 2023} & 
    % \multirow{2}{*}{{Artix-7}} &
    % {FNTT} & 
    % \multirow{2}{*}{{445}} &
    % \multirow{2}{*}{{1154}} & 
    % \multirow{2}{*}{{1031}} & 
    % \multirow{2}{*}{{2}} &
    % \multirow{2}{*}{{0}} &
    % \multirow{2}{*}{{300}} & 
    % {456} & 
    % {1.52} &
    % \multirow{2}{*}{645} & 
    % {980 (3.55$\times$\textcolor{red}{$\downarrow$})} \\ \cline{3-3} \cline{10-11} \cline{13-13}
    
    % {} & {} & {INTT} & {} & {} & {} & {} & {} & {} & {456} & {1.52} & {} & {980 (3.55$\times$\textcolor{red}{$\downarrow$})} \\ \hline 

    % \multicolumn{13}{|l|}{\textbf{PIP-NTT reporting computation time using butterfly clock cycles}} \\ \hline \hline

    %%%%%%%%%%%%%%%%%%%%%%%%%%%%%%%%%%%%
    % This Work with butterfly clock cycles
    %%%%%%%%%%%%%%%%%%%%%%%%%%%%%%%%%%%

    % \multirow{2}{*}{\textbf{PIP-NTT}} & 
    % \multirow{2}{*}{{Artix-7}} &
    % {FNTT} & 
    % \multirow{2}{*}{{341 \textcolor{blue}{$\uparrow$}}} &
    % \multirow{2}{*}{{1084 \textcolor{blue}{$\uparrow$}}} & 
    % \multirow{2}{*}{{324 \textcolor{blue}{$\uparrow$}}} & 
    % \multirow{2}{*}{{6}} &
    % \multirow{2}{*}{{2}} &
    % \multirow{2}{*}{{200}} & 
    % {520} & 
    % {2.60} &
    % \multirow{2}{*}{1341} & 
    % {3486} \\ \cline{3-3} \cline{10-11} \cline{13-13}
    
    % {} & {} & {INTT} & {} & {} & {} & {} & {} & {} & {520} & {2.60} & {} & {3486} \\ \hline  

    \end{tabular}
    \footnotesize{\\ $^1$ Slices (marked by $^{*}$) are approximated by LUTs $\times$ 0.25 + FFs $\times$ 0.125. This approximation method is taken from~\cite{SEC_calculation}. \\ 
    % $^2$ 36Kb BRAM slices. \malik{In PIP-NTT, four BRAMs are utilized, and the synthesis tool infers each BRAM size as 18KB, equivalent to two 36KB blocks.}\\ 
    $^2$ SEC = (\#BRAMs $\times$ 200) + (\#DSPs $\times$ 100) + \#Slices. Taken from~\cite{SEC_calculation, ref_for_SEC}, where one DSP and 36Kb BRAM correspond to 102.4 and 196.2 slices. 
    }
    % \vspace{-5mm}
    \label{tab:comparisons}
    %\end{center}
    \end{table*}

    \subsubsection{Comparison to designs supporting FNTT, INTT and point-wise multiplication (PWM)} 
  
    Compared to~\cite{ntt_comparison}, our~PIP-NTT achieves 3.14$\times$ and 4.02$\times$ faster FNTT and INTT computations, respectively, due to its 1.73$\times$ higher operating frequency, achieved by an 8-stage pipelined butterfly unit, which is not used in the reference design. However, the reference work is 1.33$\times$ more area-efficient in SEC, as it utilizes two BRAMs and one ROM externally, while PIP-NTT employs four $\frac{n}{4}$-sized BRAMs and two $\frac{n}{2}$-sized ROMs within the NTT core. Overall, PIP-NTT outperforms 2.34$\times$ and 3.00$\times$ in higher ATP. 

   \malik{As shown in Table~\ref{tab:comparisons}, PIP-NTT outperforms the configurable NTT core of~\cite{COHA_NTT} in SEC, achieving 3.55$\times$ and 4.57$\times$ higher ATP efficiency for FNTT and INTT, respectively. While~\cite{COHA_NTT} supports multiple $n$ and $\texttt{q}$ values, PIP-NTT is tailored for $n=256$ and $\texttt{q}=3329$. Compared to the high-performance design in~\cite{ntt_kyber_16BTF} with 16 butterfly units and 35 BRAMs, which is 6.50$\times$ and 6.34$\times$ faster, PIP-NTT—using only two butterfly units—is 9.53$\times$ more efficient in SEC and achieves 1.46$\times$ and 1.50$\times$ better ATP efficiency. In~\cite{AREA_TIME_ML_KEM_4BTU_2025}, Kyber’s full implementation includes NTT area data but lacks timing results, so ATP cannot be compared; however, PIP-NTT is 1.02$\times$ more efficient in SEC.}

    %% =====================================
    % -- Comparing designs supporting only the FNTT and INTT operations
    %% =====================================
    
    \subsubsection{Comparison to designs supporting only the FNTT and INTT operations} We compare the high-speed, state-of-the-art NTT accelerators in Section~\ref{par:SOTA_high_speed}, while other NTT-related architectures, including the area-efficient ones, are compared in Section~\ref{par:SOTA_other}.  

    %% =====================================
    % -- Comparing high-speed designs
    %% =====================================

    \paragraph{Comparison to high-speed accelerators}\label{par:SOTA_high_speed} 

    \malik{Our design targets an area-optimized accelerator and, therefore, focuses on partial parallelism for iterative NTT implementations. A like-for-like comparison of PIP-NTT with high-speed accelerators is not entirely fair because the design premises differ. For completeness, Table~\ref{tab:comparisons} shows that PIP-NTT uses fewer standalone slices, LUTs, and FFs, and achieves a lower SEC than~\cite{HyperNTT},~\cite{poly_mult_acc_for_K_2024},~\cite{K_NTT_64_BTFs_2024}, and~\cite{high_speed_NTT_kyber_2024}. In terms of ATP, the architectures in~\cite{HyperNTT} and~\cite{poly_mult_acc_for_K_2024} are more efficient due to lower cycle counts and higher operating frequency. By contrast, against the high-speed designs in~\cite{K_NTT_64_BTFs_2024} and~\cite{high_speed_NTT_kyber_2024}, PIP-NTT achieves better overall efficiency.
    }

    %% =====================================
    % -- Comparing other NTT designs
    %% =====================================

    \paragraph{Comparing other NTT-related accelerators}\label{par:SOTA_other}

    A Kyber architecture in~\cite{pipeline_ntt} employs efficient resource sharing. PIP-NTT outperforms its pipelined data path with a 1.29$\times$ higher frequency due to fine-grained butterfly pipelining, achieving a 4.55$\times$ faster computation. However, a direct comparison of area and ATP is not feasible as the reference design does not provide detailed resource utilization. Compared to a unified design of the Kyber and Dilithium NTT in~\cite{CRYPHTOR}, PIP-NTT achieves 3.03$\times$ faster computation, 1.83$\times$ higher SEC efficiency, and 5.58$\times$ better ATP performance.

    \malik{Using an optimized modular-reduction unit and the Brent–Kung method for modular addition/subtraction, the area-efficient design in~\cite{K_NTT_only_AREA_2024} uses 10 DSPs  \& 11 BRAMs, and in overall consumes $3.01\times$ more area in SEC than PIP-NTT. In our work, we reduce the area by using a multiplication-free design for $a\cdot n^{-1}~\texttt{mod q}$. The computation time and ATP cannot be compared because they are not reported in~\cite{K_NTT_only_AREA_2024}.
    } In comparison to a non-pipelined NTT accelerator of~\cite{sajjad_access_ntt}, our PIP-NTT is 1.11$\times$ more area-efficient. Moreover, the pipelined butterfly unit in PIP-NTT results in 1.62$\times$ and 1.85$\times$ faster FNTT and INTT computations, respectively, leading to 1.80$\times$ and 2.06$\times$ higher ATP efficiency. \malik{In comparison to the most recent iterative NTT design of~\cite{iter_ntt_2025_wiley}, our memory-parallelization approach in PIP-NTT for partial parallelism ensures 1.63$\times$ lower computation time, while our multiplication-free design for $a\cdot n^{-1}~\texttt{mod q}$ results 1.41$\times$ higher efficiency in ATP. 
    }
    
    For accelerators where PIP-NTT outperforms in ATP, Table~\ref{tab:comparisons} shows that~\cite{ntt_comparison} has the lowest SEC, while~\cite{ntt_kyber_16BTF} achieves the lowest computation time. Their respective average ATP values for FNTT and INTT are 8262 and 4581, while PIP-NTT achieves an average ATP of 3084. This shows that PIP-NTT is 2.67$\times$ more efficient in ATP than the most area-optimized NTT accelerator and achieves a 1.48$\times$ higher ATP performance than the most efficient high-speed NTT design. These results highlight the superior efficiency of the PIP-NTT compared to state-of-the-art implementations.

    %%==================================%%
    % Flexibility and Scalability Prospects of PIP-NTT
    %%==================================%%

    \subsection{\malik{Flexibility and Scalability Prospects of PIP-NTT}} \label{subsec:flexibility}

    \malik{
    A key strength of the proposed PIP-NTT accelerator lies in its adaptability across diverse cryptographic workloads. While initially optimized for Kyber parameters, both the core butterfly architecture and the memory-parallelization strategy are designed to scale with minimal modifications. This section outlines how PIP-NTT can be extended to support other PQC schemes, such as Dilithium, and how its memory subsystem remains robust across varying NTT configurations.
    }

    \malik{
    \subsubsection{Adaptability of PIP-NTT to other cryptographic schemes} 
    The PIP-NTT architecture, originally tailored for Kyber parameters, can be readily adapted to other PQC schemes such as Dilithium. Specifically for Kyber, transitioning from radix-2 to radix-$r$ requires only minor reordering of the butterfly’s arithmetic and \texttt{mod q} operations. To deploy PIP-NTT in Dilithium or other PQC schemes, the Kyber-specific \texttt{mod q} datapath needs to be updated to adjust the arithmetic requirements of the target scheme, and the memory/ROM width and depth must be resized according to the $\texttt{mod q}$ of the targeted algorithm.
    }
    
    \malik{
    \subsubsection{Adaptability of memory-parallelization strategy across different NTT configurations and cryptographic schemes}
    Our memory-parallelization strategy is inherently radix- and scheme-agnostic because the $\delta$-driven pairing, adjacent-coefficient processing, offset-based swapping (whether $\delta \ge n/4$ or $\delta < n/4$), and coefficient reordering remain unchanged and scale parametrically with $n$ and $\log_r n$. This also ensures the applicability of our approach in hybrid/unified designs that support multiple schemes. Only the FSM/control for address generation, stage counter, and the ROM for twiddle factors require minimal changes based on the target parameters. 
    % These lightweight modifications make the proposed architecture highly versatile for computing NTTs across various PQC algorithms and high-degree homomorphic encryption schemes.
    These lightweight modifications make the architecture broadly applicable across PQC algorithms, high-degree homomorphic encryption schemes, and hybrid multi-scheme deployments.
    }

    %%==================================%%
    % Section Conclusions
    %%==================================%%
    \section{Conclusions}\label{sec:conclusions}

    \malik{This work presents PIP-NTT, a pipelined and scalable memory-parallelized NTT accelerator evaluated on an Artix-7 FPGA for PQC KEM Kyber parameters. The combination of coarse-grained and fine-grained pipelining in the butterfly units improves the operating frequency. Moreover, using smaller distributed memories in-parallel enables partial parallelism for the iterative NTT designs without incurring memory overhead. Furthermore, a modular-addition-based approach for the rescaling step of the inverse NTT optimizes hardware resources and improves the critical path. 
    
    PIP-NTT can operate at a maximum frequency of 200 MHz, computing a single FNTT and INTT in 2.60$\mu s$. It demonstrates 2.67$\times$ and 1.48$\times$ higher efficiency in average ATP compared to the most area-efficient and high-speed NTT accelerators reported in the literature. The PIP-NTT design is not limited to Kyber; its radix- and scheme-agnostic architecture makes it a promising candidate for broader adoption in future lattice-based cryptographic accelerators.
    }
    
    \bibliographystyle{IEEEtran}
    \vspace{-0.5mm}
    \bibliography{references}% common bib file

    \input{biography}

\vfill

\end{document}

%% file: biography.tex
\vspace{-10mm}

\begin{IEEEbiography}[{\includegraphics[width=1in,height=1.25in,clip,keepaspectratio]{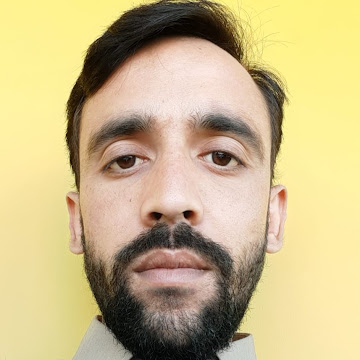}}]{Malik Imran} 
        is a Research Fellow at the Centre for Secure Information Technologies (CSIT), Queen’s University Belfast, Northern Ireland. He received his Ph.D. in Information and Communication Technologies from Tallinn University of Technology (TalTech), Estonia, in 2023. He received M.S. and B.S. degrees in Telecommunication and Networks and Computer Engineering from Pakistan in 2015 and 2011, respectively. His research interests focus on secure-by-design hardware accelerators for cryptographic primitives, with emphasis on post-quantum cryptography, homomorphic encryption and privacy computing. 
    \end{IEEEbiography}

    \vspace{-15mm}

    \begin{IEEEbiography}[{\includegraphics[width=1in,height=1.25in,clip,keepaspectratio]{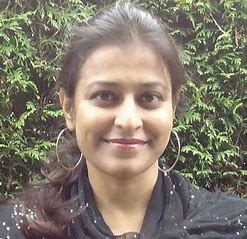}}]{Ayesha Khalid} 
        (M'2017--SM'2022) is a Senior Lecturer with Queen's University of Belfast, Belfast, U.K. She received her the B.E. degree in Computer Systems Engineering from the National University of Sciences and Technology (NUST), Pakistan, and the M.S. degree in Electrical Engineering from the Center for Advanced Studies in Engineering (CASE), University of Engineering and Technology, UET-Taxila. Her research interests include lattice-based cryptography, embedded systems security, side channel attacks, and cryptographic hardware. She was a recipient of the DAAD Scholarship Award for her Ph.D. studies at RWTH Aachen, Germany.
    \end{IEEEbiography}

    \vspace{-10mm}

    \begin{IEEEbiography}[{\includegraphics[width=1in,height=1.25in,clip,keepaspectratio]{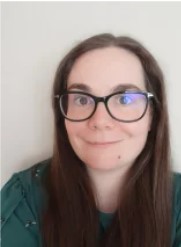}}]{Ciara Rafferty} 
        (M'14, SM'24) is a Senior Lecturer at the Centre for Secure Information Technologies (CSIT) in Queen’s University Belfast, Northern Ireland. She received her BSc degree in Mathematics with Extended Studies in Germany from Queen's University Belfast in 2011 and the PhD degree in Electrical and Electronic Engineering from Queen's University Belfast in data security in 2015. Her research interests include post-quantum cryptography, applied cryptography, privacy enhancing technologies and optimised hardware architectures of cryptographic primitives.
    \end{IEEEbiography}

    \vspace{-10mm}

    \begin{IEEEbiography}[{\includegraphics[width=1in,height=1.25in,clip,keepaspectratio]{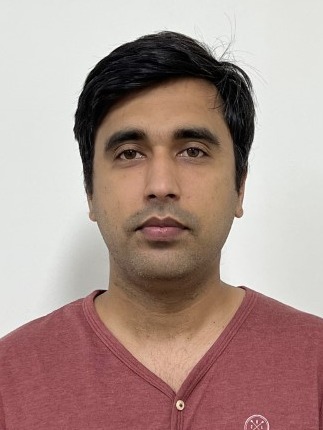}}]{Safiullah Khan} 
        received his Ph.D. degree in computer engineering from Gachon University, South Korea, in 2023. He was a Project Engineer with the Research and Development Department, National Radio and Telecommunication Corporation, Haripur, Pakistan, for two years. He also remained a Research Fellow with the Centre for Secure Information Technologies (CSIT), Institute of Electronics, Communications and Information Technology (ECIT), Queen’s University, Belfast, U.K. Currently, he is a Lecturer with the Department of Computing and Mathematics, Manchester Metropolitan University, Manchester, U.K. His research interests include efficient hardware implementations of cryptographic protocols, post-quantum cryptography, and homomorphic encryption.
    \end{IEEEbiography}

    \vspace{-10mm}

    \begin{IEEEbiography}[{\includegraphics[width=1in,height=1.25in,clip,keepaspectratio]{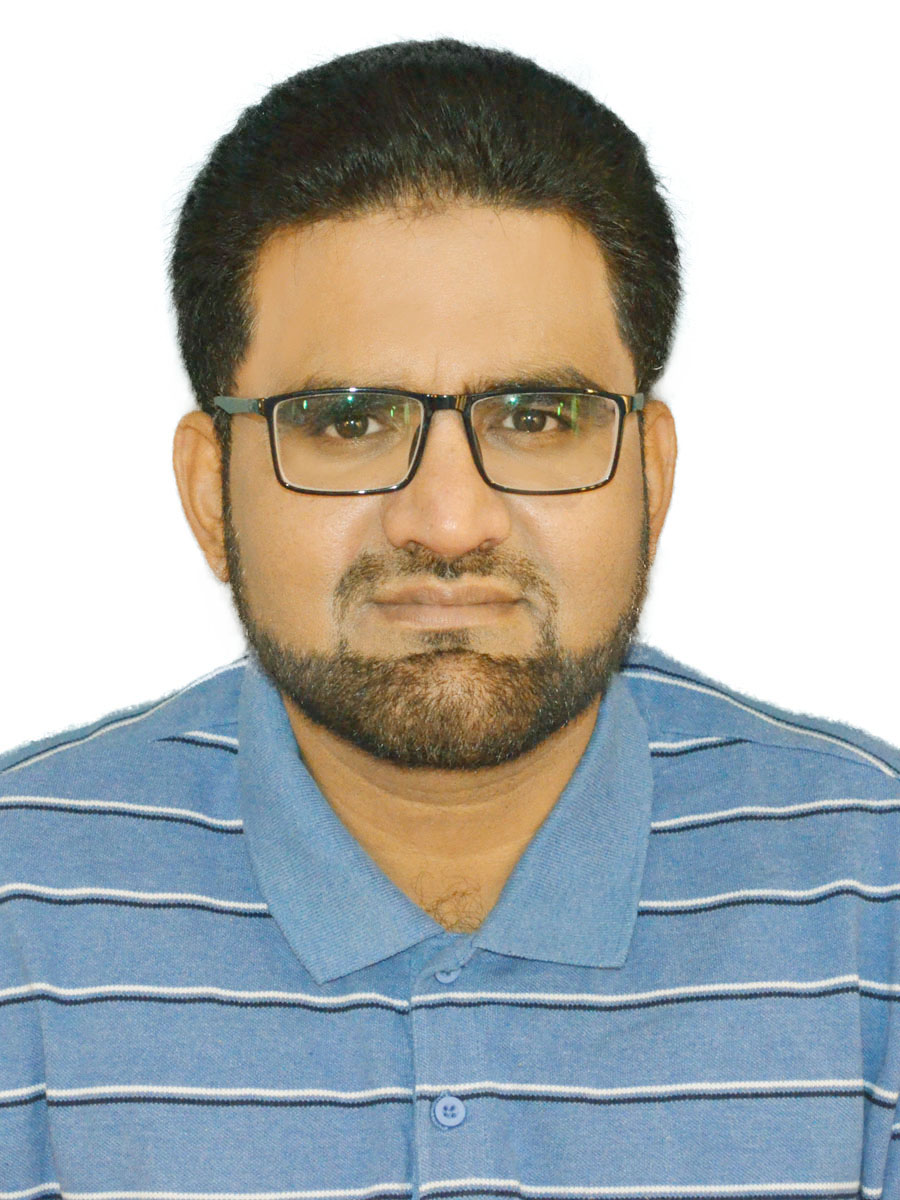}}]{Muhammad Rashid} 
        is a distinguished professor at Umm Al-Qura University, renowned for his expertise in embedded systems. With over 24 years of academic and industrial experience, he has made significant contributions to the fields of electrical and computer engineering. He has published approximately 140 articles in prestigious journals and conference proceedings. He holds a PhD in Information and Communication from Université de Bretagne Occidentale and a Master's in Embedded Systems from Université de Nice Sophia Antipolis. Throughout his career, Dr. Rashid has held key positions, including Manager of Research and Development at Advanced Engineering Research Organization.
    \end{IEEEbiography}

    \vspace{-10mm}

    \begin{IEEEbiography}[{\includegraphics[width=1in,height=1.25in,clip,keepaspectratio]{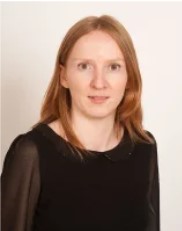}}]{Maire O'Neill} 
        (M’03-SM’11) received the M.Eng. degree with distinction and the PhD degree in electrical and electronic engineering from Queen’s University Belfast, U.K., in 1999 and 2002, respectively. She is a Chair of Information Security at Queen’s and holds an EPSRC Leadership fellowship to conduct research into next generation data security architectures. She previously held a UK Royal Academy of Engineering research fellowship from 2003 to 2008. She has authored two research books and has more than 100 peer-reviewed conference and journal publications. Her research interests include hardware cryptographic architectures, side channel analysis, physical unclonable functions and post-quantum cryptography. In 2014 she was awarded a Royal Academy of Engineering Silver Medal, which recognises outstanding personal contribution by an early or mid-career engineer that has resulted in successful market exploitation.
    \end{IEEEbiography}